\documentclass[twocolumn,tighten,twocolappendix]{aastex63}

\usepackage{amsmath}
\usepackage{enumerate}

\usepackage[shortlabels]{enumitem}
\setlist[enumerate]{nosep,nolistsep}
\setlist[itemize]{nosep,leftmargin=*}

\RequirePackage{xcolor,colortbl}
\newcommand{\kms}[1]{{km\,s$^{-1}$#1}}
\newcommand{\ms}[1]{{m\,s$^{-1}$#1}}
\newcommand{\ARA}[1]{AR\,10930}
\newcommand{\ARB}[1]{AR\,11967}
\newcommand{\animationone}[1]{\href{https://www.dropbox.com/s/zymo1dc52vz8adq/Animation1_CEF-1_and_CEF-2.mp4?dl=0}{Animation~1}}
\newcommand{\animationtwo}[1]{\href{https://www.dropbox.com/s/979ebbylxo16kvh/Animation2_CEF-3.mp4?dl=0}{Animation~2}}

\definecolor{darkgreen}{rgb}{0.0, 0.4, 0.0}

\usepackage{txfonts}
\usepackage{booktabs}

\graphicspath{{figures2expulsionCEFs/}}

\shorttitle{Expulsion of Counter Evershed Flows}
\shortauthors{{Castellanos~Dur\'{a}n}, Korpi-Lagg \& Solanki}

\begin{document} 

\title{Expulsion of counter Evershed flows from sunspot penumbrae}
\correspondingauthor{J.~S. {Castellanos~Dur\'an}}
\email{castellanos@mps.mpg.de}

\author[0000-0003-4319-2009]{J.~S. {Castellanos~Dur\'an}}
\affiliation{Max Planck Institute for Solar System Research, Justus-von-Liebig-Weg 3, D-37077 G\"ottingen, Germany}
\affiliation{Georg-August-Universit\"at G\"ottingen, Friedrich-Hund-Platz 1, D-37077. G\"ottingen, Germany}

\author[0000-0003-1459-7074]{A. Korpi-Lagg}
\affiliation{Max Planck Institute for Solar System Research, Justus-von-Liebig-Weg 3, D-37077 G\"ottingen, Germany}
\affiliation{Department of Computer Science, Aalto University, PO Box 15400, FI-00076 Aalto, Finland}

\author[0000-0002-3418-8449]{S.~K. Solanki}
\affiliation{Max Planck Institute for Solar System Research, Justus-von-Liebig-Weg 3, D-37077 G\"ottingen, Germany}
\affiliation{School of Space Research, Kyung Hee University, Yongin, 446-101 Gyeonggi, Republic of Korea}

\begin{abstract}
   In addition to the Evershed flow directed from the umbra towards the outer boundary of the sunspot, under special circumstances, a counter Evershed flow (CEF) in the opposite direction also occurs. We aim to characterize the proper motions and evolution of three CEFs observed by the Solar Optical Telescope onboard the Japanese Hinode spacecraft and the Helioseismic and Magnetic Imager onboard the Solar Dynamics Observatory. We use state-of-the-art inversions of the radiative transfer equation of polarized light applied to spectropolarimetric observations of the \ion{Fe}{1} line pair around 630\,nm. The three CEFs appeared within the penumbra. Two of the CEF structures, as part of their decay process, were found to move radially outwards through the penumbra parallel to the penumbral filaments with speeds, deduced from their proper motions, ranging between 65 and 117\,m\,s$^{-1}$. In these two cases, a new spot appeared in the moat of the main sunspot after the CEFs reached the outer part of the penumbra. Meanwhile, the CEFs moved away from the umbra, and their magnetic field strengths decreased. The expulsion of these two CEFs seems to be related to the normal Evershed flow. The third CEF appeared to be dragged by the rotation of a satellite spot. Chromospheric brightenings were found to be associated with the CEFs, and those CEFs that reached the umbra-penumbra boundary showed enhanced chromospheric activity. The two CEFs, for which line-of-sight velocity maps were available during their formation phase, appear as intrusions into the penumbra. They may be associated with magnetic flux emergence.

\end{abstract}
   \keywords{Sunspot groups (1651); Solar photosphere (1518); Solar magnetic fields (1503);  Solar active region velocity fields (1976); Solar chromosphere (1479); Solar flares (1496).}

\section{Introduction}

Evershed flows are characteristic outflows observed in the penumbrae of sunspots \citep{Evershed1909} with typically subsonic velocities of $\sim$1-3\,\kms{} in the body of the filament \citep[e.g.,][]{Schlichenmaier2000A&A...EFobs,Strecker2022A&A...EF} and 5-10\,\kms{} at the endpoints \citep[e.g.][]{Tiwari2013}. The characteristic filamentary structure of penumbrae observed in continuum images is the result of the interaction between buoyant convective cells rising from the solar interior and inclined magnetic field \citep[see][for reviews]{Solanki2003, Borrero2011LRSP}. The \textit{normal} Evershed flows transport plasma radially\footnote{The term `radial' is referring to the direction along the solar surface away from the center of the sunspot.} outwards along the penumbral filaments \citep[= intraspines; e.g.,][]{Lites1993ApJSpinesInPenumbra, Jurcak2007PASJ,Borrero&solanki2008ApJ}. In the last decade, penumbral regions with the opposite direction of the flow at photospheric layers, but otherwise indistinguishable in the continuum images, were observed \citep{Kleint2012ApJ...Cflare2CEF,Kleint2013ApJ, Louis2014A&A...CEF, Siu-Tapia2017A&A,CastellanosDuran2021...rareCEFs}. The new type of penumbral flow was named counter Evershed flow (CEF) to distinguish it from the distinct chromospheric inverse Evershed flow \citep[e.g.,][]{StJohn1911ApJb....IEF, StJohn1911ApJa....IEF, Choudhary2018ApJ...IEF, Beck2020ApJ...891..IEF}. CEFs have also been observed in ideal magnetohydrodynamic simulations \citep[MHD;][]{Siu-Tapia2018ApJ}.

\citet{Louis2014A&A...CEF} did  one of the first specific analyses of a CEF. They reported a maximum line-of-sight velocity of 1.6\,\kms{}, an area of 5.2 arcsec$^2$ ($\sim$2.6\,Mm\,$^2$), and a lifetime of 1\,h for the single event they studied. These authors associated these flows with the evolution of the sunspot, which fragmented two days after the analyzed observations. \citet{Siu-Tapia2017A&A} found that the global properties inside a CEF, such as temperature, magnetic field strength ($B$), and the line-of-sight velocity ($v_{{\rm LOS}}$) vary with height similarly to the properties in the parts of the penumbra displaying the normal Evershed flow. Nonetheless, at the umbra-penumbra boundary, magnetic fields with strengths of up to 8.2\,kG  and $v_{\rm LOS}\gtrsim15$\,\kms{} at optical depth unity ($\tau=1$) were reported \citep{Siutapia2019}. 

Recently, \citet{CastellanosDuran2021...rareCEFs} reported that CEFs appear ubiquitously  in all types of sunspots. These authors found almost $\sim$400 CEFs in their survey and documented different types of CEFs. In particular, they distinguished between those that appear in  penumbrae bordering on regular umbrae, and those CEFs that are linked to light bridges.

When analyzing the different contributions in the momentum equation inside a simulated box from an MHD simulation, \citet{Siu-Tapia2018ApJ} confirmed that the normal Evershed flow is a result of the overturning of the hot material coming from the solar interior in the presence of an inclined magnetic field \citep{Rempe2009Sci, Rempel2011ApJ...Largescaleflows}. The CEFs in the simulations are, according to \citet{Siu-Tapia2018ApJ}, compatible with siphon flows, however. Penumbral siphon flows result from asymmetric heating inside the flux tube that produces the required difference in gas pressure to drive material along the arched magnetic tubes \citep{Thomas1993ApJ, Montesinos1997Natur...EF}, although, in CEFs, the siphon flows point in the opposite direction to the normal Evershed flow.

Although the maintenance of CEFs during their steady phase, at least in the MHD simulations, can be explained by the siphon flow mechanism, it remains unclear, what process leads to the formation of the opposite direction to the Evershed flow.Possible candidates identified by observers are
flux emergence \citep[e.g.,][]{Louis2014A&A...CEF, Louis2020...cefs}, the adhesion of the penumbra from another spot after two spots merge \citep{Siu-Tapia2017A&A}, as well as
the association of granular and filamentary light bridges and CEFs \citep{CastellanosDuran2021...rareCEFs}.

The evolution over time of CEFs is still barely known \citep[cf.][]{Louis2020...cefs}. In contrast, the motion of another type of magnetic feature inside sunspot penumbrae has been the topic of numerous studies. The expulsion of so-called `sea-serpent' magnetic fields lines was observed mainly in the plage surrounding the sunspot, but also in the penumbra itself  \citep{SainzDalda2008A&AL...sea-serpent}. These small, bipolar features have a filamentary structure, their length ranges between 2\arcsec and 5\arcsec and with a mean width of 1.5\arcsec. They appeared in the mid-penumbra and are expelled radially outwards with velocities ranging from 0.3--0.7\,\kms{}. Their lifetime ranges from 30\,min up to 7\,h. After the expulsion, these structures continue to travel in the moat up to 3--6\arcsec away from the penumbral boundary into the surrounding plage region. The same authors suggested that these bipolar structures are moving U-loops driven by the Evershed flow and are the precursors of moving magnetic features \citep[MMF;][]{Harvey1973SoPh...MMF, Zhang2003A&A...MMF, SainzDalda2005ApJ...MMF, Zhang2007ApJL...CEF}. Also, the so-called Evershed clouds prominent in proper motion studies, have been related to MMFs \citep{CabreraSolana2006ApJL...EFclouds}.

The moat flow is a horizontal radially outward orientated flow starting from the outer part of the penumbra and connecting the penumbral filaments with the quiet Sun \citep[e.g.,][]{Sheeley1969SoPh...MMF, VargasDominguez2007ApJL...moatflow,VargasDominguez2008ApJ...moatflow,Strecker2018A&A...moatflow}. The typical velocity of the moat outflow ranges between 0.8 and 1.4\,\kms{} and it vanishes abruptly at a distance of similar length as the penumbra away from the outer penumbral boundary \citep{Sobotka2007A&A...moat,LoehnerBoettcher2013A&A...moatflow}.

In this work, we study the thermal and velocity conditions, magnetic field structure, and the temporal evolution of three CEFs observed in \ARA{} (solar cycle 23) and \ARB{} (solar cycle 24). Two of these CEFs are seen to be expelled radially outwards beyond the outer boundary into the moat of the main sunspot within the Active Region (AR).  The host sunspots of these CEFs have been widely studied not only due to their peculiar flows, but also because they belong to ARs that harbored superstrong magnetic fields \citep{Siu-Tapia2017A&A, Okamoto2018ApJ, Siutapia2019, CastellanosDuran2020}, and  \ARA{} hosted four large X-class flares. These solar flares are among the most studied and modeled X-class flares of solar cycle 23 \citep[e.g.,][and references therein]{Wang2008ApJAR10930, Schrijver2008ApJAR10930, Gosain2009ApJ, Fan2011ApJ...740...Xclassflare2CEF, Fan2016ApJ, Wang2022...XflareAR10930}.

In this study, we aim to characterize the temporal evolution of three CEFs. In particular, we analyze their appearance, evolution and expulsion, and describe the new magnetic configuration after their expulsion. In addition, we discuss the chromospheric response to the presence of CEFs. 

This article is arranged as follows: Section~\ref{sec:expulsionCEFsMethods} introduces the data and the applied inversion method to retrieve the physical conditions within the CEFs from spectropolarimetric data. Sections~\ref{sec:cefAR10930} and \ref{sec:cefAR11967} describe the properties of the three studied CEFs. The appearance and expulsion of CEFs are presented in Sections~\ref{sec:expulsionCEF:apperance} and \ref{sec:expulsion}. Section~\ref{sec:expulsionCEFs:satelliteSpot} illustrates the evolution of the magnetic regions that are left after the expulsion of CEFs. In Section~\ref{sec:magdistriExpulsion}, we describe the variation of $B$ and $v_{\rm LOS}$ within the CEFs. The chromospheric response to the presence of CEFs is presented in Section~\ref{sec:expulsitionCEF:Chrosmophere}. In Section~\ref{sec:expulsionCEFs:discussion}, we discuss our results and we conclude in Section~\ref{sec:expulsionCEFs:conclusion}.

 \begin{figure*}[phtb]
     \centering
     \includegraphics[width=.72\textwidth]{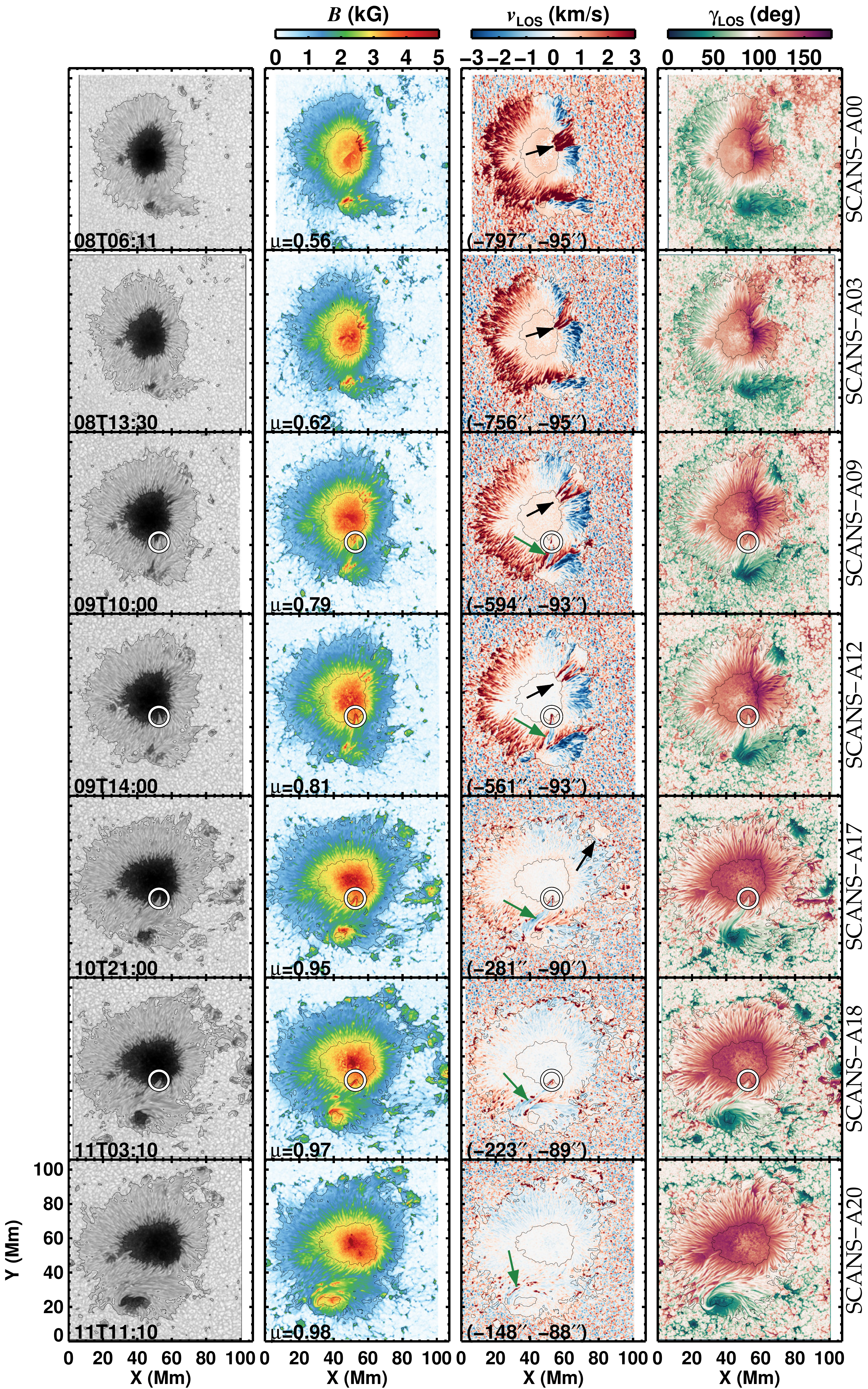}
      \caption{Temporal evolution of \ARA{} as observed by Hinode/SOT-SP. Time runs from top to bottom. Columns are the temperature, the magnetic field strength $B$, $ v_{{\rm LOS}}$ and the inclination of the magnetic field in the line-of-sight. Contours show the inner and outer penumbra boundaries. The black and green arrows mark CEF-1 and CEF-2, respectively. White circles shown on the four bottom rows mark an intrusion into the umbra associated with CEF-2. See Figure~\ref{fig:vlosAR11967_tip} for a zoom-in of this intrusion. See also \animationone{} part of the online material.}\label{fig:vlosAR10930}
 \end{figure*}

\section{Observations and methods}\label{sec:expulsionCEFsMethods}

\subsection{Data}
We observed two sunspot groups from two different solar cycles. The sunspot group \ARA{} was followed for 8 days starting on 2006 December 8, and the sunspot group \ARB{} for 6 days starting from 2014 February 1.  We analyzed spectropolarimetric observations taken by the Japanese Hinode mission launched in 2006 \citep{Kosugi2007}. The Spectro-Polarimeter \citep[SP;][]{Ichimoto2008SoPh} aboard Hinode measures the four Stokes parameters $(I, Q, U, V)$ of the \ion{Fe}{1} line pair around 6302\,\AA{}, with a spectral sampling of 21.5\,m\AA{}. We analyzed 42 scans of  \ARA{} and 32 of \ARB{} (hereafter \texttt{SCANS-A(00-41)} and \texttt{SCANS-B(00-31)}, respectively).  The spatial sampling along the slit and scan direction can be either 0\farcs16 (normal mode) or 0\farcs32 (fast mode) depending on the observing mode.  Data were reduced using the nominal Hinode/SOT-SP pipeline \texttt{sp\_prep} \citep{Lites2013SoPh}.  We also analyzed all the available photospheric $G$-band filtergrams and the chromospheric \ion{Ca}{2}\,H images taken by Hinode/SOT-BFI \citep{Tsuneta2008}, and the Stokes $V$ maps from Hinode/SOT-NFI \citep{Tsuneta2008} recorded in the intervals 2006 December 6 to 15 and 2014 February 1 to 6.

\begin{figure*}[phtb]
     \centering
     \includegraphics[width=.63\textwidth]{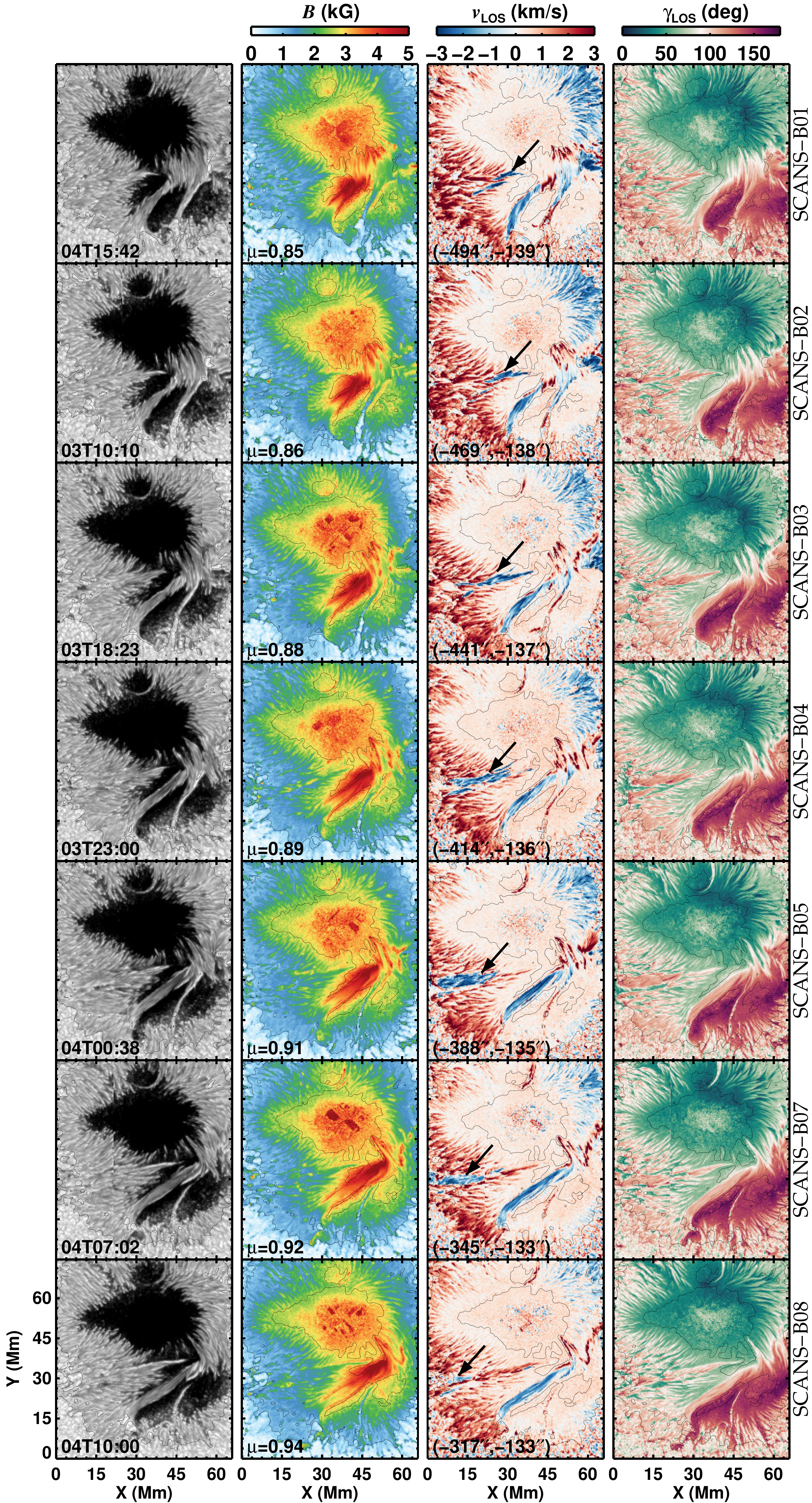}
    \caption{Same layout as Figure~\ref{fig:vlosAR10930} for \ARB{}. Black arrows indicate the location of CEF-3. See also \animationtwo{}, which is part of the online material. }\label{fig:vlosAR11967}
 \end{figure*}

We use the following nomenclature throughout the paper: Letters \texttt{A} and \texttt{B} are used to differentiate the Hinode/SOT-SP \texttt{SCANS} of \ARA{} and \ARB{}, respectively. Notice that for \ARA{}, the Hinode/SOT-SP scans covered the entire sunspot group; however, for \ARB{}, many of the Hinode/SOT-SP scans focused only on the eastern group of sunspots. We restrict our analysis to the eastern group containing one of the CEFs, accounting for approximately $\sim$1/3 of the total sunspot area within \ARB{}. The left columns in Figures~\ref{fig:vlosAR10930} and \ref{fig:vlosAR11967} show a continuum image each of parts of \ARA{} and \ARB{} (see following sections for details). We use numbers 1 to 3 to mark the three CEFs analyzed in detail in this study.

In addition, we used data from the Solar Dynamic Observatory \citep[SDO;][]{Pesnell2012} taken by the Helioseismic and Magnetic Imager \cite[HMI;][]{Scherrer2012,Schou2012}. We analyzed the continuum intensity, Dopplergrams ($v_{{\rm LOS}}$), and magnetograms ($B_{{\rm LOS}}$) obtained at a spatial resolution of 1\arcsec{}. Two intervals were used with each of them having different cadences and fields-of-view (FOV). The first interval covered the entire passage of \ARB{} over the solar disk from 2014 January 28 at 20:00\,UT to February 8 at 20:00\,UT at a cadence of 12 minutes. The second interval started on 2014 February 1 at 04:00\,UT and lasted until February 2 at 12:00\,UT with the data taken at a cadence of 45 seconds. The FOV of the first dataset was cropped to cover the entire \ARB{}, whilst the second FOV was cropped to cover the same region as observed by Hinode/SOT-SP, but extended to include the eastern moat of the main sunspot of \ARB{} (see \animationtwo{}). Continuum maps were corrected for limb darkening following \citet{CastellanosDuran2020...blow2wl}.

\subsection{Inversion scheme}
To extract the physical information encoded in the Stokes profiles, we used the Stokes Profiles Inversion-O-Routines (SPINOR) inversion code \citep{Frutiger2000}. SPINOR builds on the STOkes PROfiles routines (STOPRO) that solve the radiative transfer equations for polarized light \citep{Solanki1987PhDT}. In the \textit{traditional} scheme, SPINOR (as well as other inversion codes commonly used in solar physics; e.g., the Stokes Inversion based on Response functions code \citep[SIR;][]{RuizCobo1992ApJ...SIR}, the He-Line Information extractor$^+$ code \citep[HeLIx$^+$;][]{Lagg2004,Lagg2009}, the HAnle and ZEeman Light code \citep[HAZEL;][]{AsensioRamos2008}, the Spectropolarimetic NLTE Analytically Powered Inversion code \citep[SNAPI;][]{Milic2018A&A...SNAPI}), invert each pixel $(x,y)[I(\lambda),Q(\lambda),U(\lambda),V(\lambda)]$ within the FOV independently. However, these pixels are spatially coupled due to the action of the point spread function (PSF) of the telescope. Recently, the spatially-coupled
concept has been extended into the STockholm inversion Code \citep[STIC;][]{delaCruzRodriguez2019A&A...STIC} to account for simultaneous observations
taken by different instruments with intrinsically different
PSFs \citep{delaCruzRodriguez2019AA...coupled}.

For the data analyzed here, the pupil of Hinode/SOT with its central obscuration and the triangular spider produces a complex radially non-symmetric PSF \citep[\citet{Danilovic2008A&AL...contrast}; cf. Figure~10 in][]{vanNoort2012A&A}. This complex PSF couples the information of neighboring pixels and needs to be taken into account when analyzing Hinode/SOT-SP observations. This was achieved when \citet{vanNoort2012A&A} developed the spatially coupled scheme for inversions and implemented it into SPINOR (hereafter spatially coupled inversion) that treated both, the spectropolarimetric information and the inherent spatial degradation caused by the spatial PSF. This technique was improved by showing that improved results are obtained by applying it to finer, i.e. interpolated spatial pixels \citep{vanNoort2013A&A}.

 \begin{figure*}[phtb]
     \centering
     \includegraphics[width=.65\textwidth]{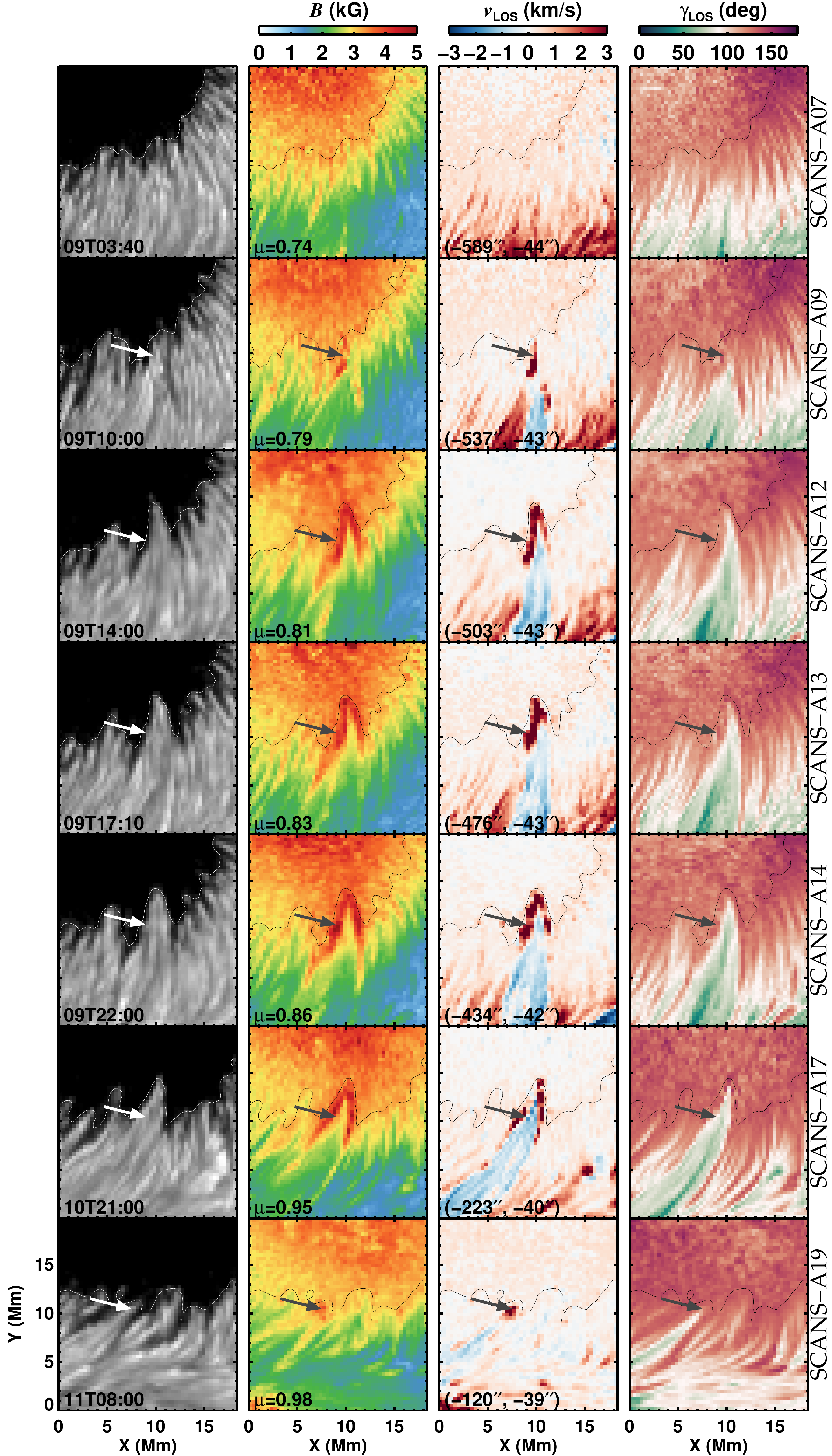}
    \caption{Zoom-in into the northern arrowhead-shaped region referred to as the tip of CEF-2 in the main text (marked by the arrows). Time runs from top to bottom. Columns are the temperature, the magnetic field strength $B$, $v_{{\rm LOS}}$ and the inclination of the magnetic field relative to the line-of-sight. }\label{fig:vlosAR11967_tip}
 \end{figure*}

The spatially coupled inversions allowed to obtain excellent fits to the observed Stokes profiles, while keeping a single depth-depended atmospheric model when fitting different photospheric features \citep[see e.g.,][]{vanNoort2013A&A, Tiwari2015A&A...penumbra, CastellanosDuran2022...Phd}. The spatially coupled inversions of the Hinode/SOT-SP observations were carried out with a depth-stratified atmosphere at three-node positions for the temperature, magnetic field strength, inclination and azimuth, $v_{\rm LOS}$, and a constant value for micro-turbulence that accounts for the broadening of the spectral lines by unresolved turbulent motions. The spectral PSF is taken into account by convolving the synthetic spectra with the instrumental profile representing the spectral resolution of Hinode/SOT-SP \citep{vanNoort2012A&A}. The node positions were placed at $\log\tau=(0,-0.8,-2.0)$  for \ARA{} following \citet{Siu-Tapia2017A&A}, and at $\log\tau=(0,-0.8,-2.3)$ for \ARB{}. Maps of the retrieved atmospheric conditions for these two sunspot groups are presented in Sections \ref{sec:cefAR10930} and  \ref{sec:cefAR11967}, as well as some examples of fits to the observed Stokes profiles. 

When the spatial PSF of the optical system is known, the spatially coupled inversions can be used to estimate atmospheric conditions up to the telescope's diffraction limit. We upsampled the data by a factor of two before running the spatially coupled inversions to fit substructures that are below the spatial resolution of the telescope as recommended by \citet{vanNoort2013A&A}. After final convergence of the spatially coupled inversion, we downsampled the retrieved atmospheric conditions and the best-fit profiles to the original sampling. Data upsampling and downsampling were performed in Fourier space.

Several Hinode/SOT-SP scans of all the CEFs analyzed in this work were taken at $\mu-$values larger than 0.8, allowing us to determine their polarity with reasonable accuracy without transforming the magnetic field into the local reference frame. Examples of observed Stokes profiles and their fits obtained with spatially coupled inversions are shown in Figure~\ref{fig:strongfields}. These profiles were chosen to display that even highly complex Stokes profiles are well modelled with our inversion scheme.

\begin{figure*}[tbhp]
\begin{center}
 \includegraphics[width=.49\textwidth]{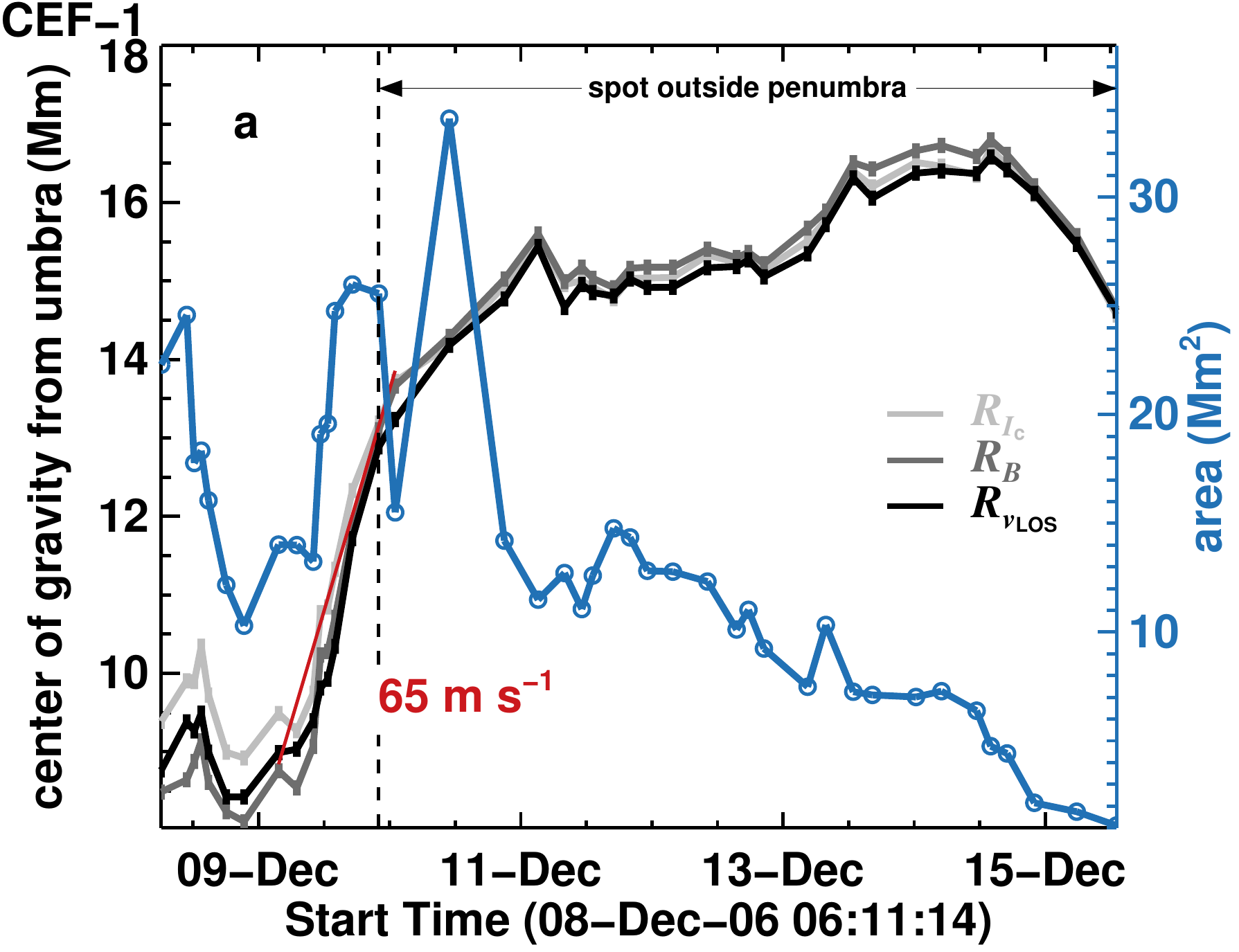}
 \includegraphics[width=.49\textwidth]{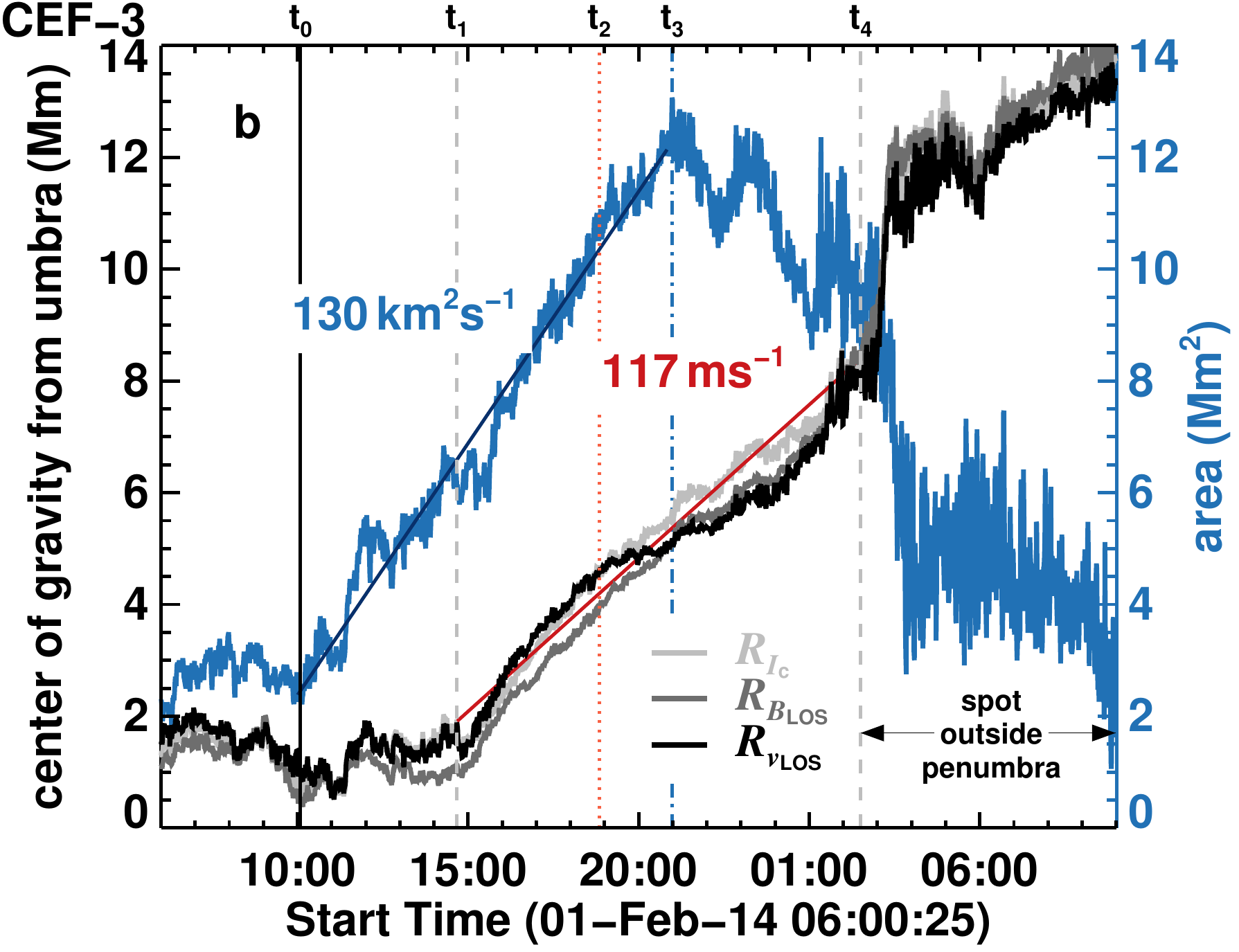}
 \caption{Temporal evolution of the center of gravity of $v_{\rm LOS}$ ($R_{v_{\rm LOS}}$; black line), magnetic field strength ($R_B$; dark gray line), and brightness ($R_{I_{\rm c}}$; light gray line), as well as the area (blue line; right axis) of CEF-1 (panel (a)) and CEF-3 (panel (b)). The vertical line on the left panel represents the time when CEF-1 is totally expelled from the penumbra of the main sunspot of \ARA{}. After this time, panel (a) shows the location of the centers of gravity and area of the  spot that formed at the location of where CEF-1 ended into the moat of \ARA{}. The vertical lines on panel (b) mark the times when CEF-3 started to grow ($t_0$, vertical solid line), when CEF-3 started to be expelled ($t_1$, vertical gray dashed line), when the LOS magnetic field and velocity had their maximum ($t_2$, vertical red dotted line), when the maximum area was reached ($t_3$, vertical blue dash-dotted line), and when CEF-3 was totally expelled from the penumbra into the moat of \ARB{} ($t_4$, vertical grey dashed line; see main text for details).  
 }\label{fig:growth}
 \end{center}
 \end{figure*}

\section{Results}

\subsection{CEFs in \ARA{}}\label{sec:cefAR10930}

The $\delta$-sunspot group \ARA{} contains two large colliding sunspots of opposite polarity, with the southern spot rotating rapidly counterclockwise. This active region hosted two CEFs, both in the penumbra of the main sunspot located in the north of \ARA{}. The complexity and rotation of the sunspots within \ARA{} influence the evolution of the CEFs that it harbored (see below).

The first CEF (CEF-1) was observed on the north-west part of this sunspot and remained within the penumbra for 17 Hinode/SOT-SP scans recorded between 2006 December 8 at 06:11\,UT (\texttt{SCANS-A00}, $\mu=0.56$) and  2006 December 10 at 21:00\,UT (\texttt{SCANS-A16}, $\mu=0.92$). CEF-1 appeared as a red-shifted region within the center-side penumbra surrounded by the normal Evershed flow, which appeared blue-shifted when \ARA{} was located on the eastern hemisphere.  

The second CEF (CEF-2) emerged on 2006 December 9 at 07:00\,UT (\texttt{SCANS-A08}, $\mu=0.76$) and completely vanished on 2006 December 11 at 11:10\,UT (\texttt{SCANS-A20}, $\mu=0.98$) before \ARA{} crossed the central meridional. CEF-2 appeared as an elongated, blue-shifted penumbral region enclosed by normal penumbra on the limb-side (i.e., the normal Evershed flow in that part of the penumbra was red-shifted). CEF-2 was located on the south-side of \ARA{}. CEF-2 connected the main umbra of \ARA{} and a smaller umbra with opposite magnetic polarity.  CEF-2 appeared like a normal Evershed flow, but oriented from the smaller umbra towards the bigger one, while on both sides of the CEF-2 the Evershed flow was dominated by the main umbra (which would be CEFs when viewed from the small umbra). This example shows the difficulty of distinguishing between the normal Evershed flow and a CEF in more complex ARs. 

Figure~\ref{fig:vlosAR10930} displays the temporal evolution of both, CEF-1 and CEF-2. Columns display from left to right the temperature, $B$, $v_{\rm LOS}$ and $\gamma_{\rm LOS}$, all at the middle node. 

The magnetic configurations of CEF-1 and -2 were very different. CEF-1 had the same polarity as the main spot in \ARA{} close to the umbra-penumbra boundary and opposite in outer penumbra. CEF-2 had opposite polarity to the surrounding penumbrae. CEF-1 covered an area starting from the umbra-penumbra boundary to the quiet Sun. CEF-2 appeared as a thin elongated filamentary structure that grew until it formed a bridge between the main north positive umbra and the growing south negative umbra. To better display the temporal evolution of the CEF-1 and -2, we co-aligned the Hinode/SOT-SP scans with each other and present them as \animationone{} among the online material.

\subsection{CEF in \ARB{}}\label{sec:cefAR11967}

Active region 11967 was one of the largest and most complex sunspot groups of solar cycle 24. We tracked \ARB{} for 11.1 days. During this period 19 CEFs were found at different parts of the sunspots belonging to this intricate active region. In this work, we focus only on one of these CEFs, which was co-observed by Hinode/SOT-SP. Hereafter we refer to this CEF as CEF-3 (Figure~\ref{fig:vlosAR11967}). CEF-3 was observed when \ARB{} was on the eastern hemisphere and it emerged as an intrusion in the penumbra with opposite polarity. CEF-3 was present in 9 out of 11 scans taken by Hinode/SOT-SP between 2014 February 1 at 10:42\,UT (\texttt{SCANS-B00}, $\mu=0.83$) and 2014 February 2 at 10:20\,UT (\texttt{SCANS-B10}, $\mu=0.96$). CEF-3 first appeared as two elongated penumbral filaments that grew and later merged (Figure~\ref{fig:vlosAR11967}, \texttt{SCANS-B00} to \texttt{B02}). It had opposite magnetic polarity compared to the surrounding penumbra and the umbra in its vicinity. CEF-3 expanded until it filled the entire length of the penumbra before it got expelled. \animationtwo{} showing the temporal evolution of CEF-3 as seen by SDO/HMI is available as online material.

In \ARB{} there is another elongated blue-shifted region in the south-west of CEF-3 (see Figure~\ref{fig:vlosAR11967}, \texttt{SCANS-B08} at (40,\,20)\,Mm). This region is a widely studied bipolar light bridge \citep{Okamoto2018ApJ, CastellanosDuran2020} that separates opposite polarity umbrae. Bipolar light bridges usually harbor bi-directional flows, which can be identified by velocities of alternating sign
\citep{CastellanosDuran2022...Phd}. Consequently, the direction of flows inside these regions cannot be classified as either normal or counter Evershed flows.

\subsection{Appearance of the CEFs}\label{sec:expulsionCEF:apperance}

Unfortunately there are no Hinode/SOT data during the appearance phase of CEF-1.  For CEF-2 and CEF-3, we could follow their entire formation process. These two CEFs appeared as intrusions inside a fully formed penumbra without any merging with an external magnetic structure (see Figures~\ref{fig:vlosAR10930} and \ref{fig:vlosAR11967}), resembling the emergence of new magnetic flux at the solar surface. This appearance process of CEF-2 and CEF-3 is better seen in \animationone{} and \animationtwo{}.

In addition, during the appearance phase of CEF-2, the northern edge of the penumbral filament that harbored CEF-2 showed a fairly distinctive behavior. As time progresses, it developed into an arrowhead-shaped intrusion of the penumbra towards the umbra. When the intrusion was fully formed, the umbra-penumbra boundary shifted by $\sim$5\,Mm towards the inner umbra. This region is encircled in Figure~\ref{fig:vlosAR10930}, centered at (56, 56)\,Mm. Figure~\ref{fig:vlosAR11967_tip} shows a zoom-in into this intrusion, revealing an enhanced $B$ at its edges.  
The flow at the tip of the intrusion is of opposite direction to CEF-2 but has the same direction of the normal Evershed flow at that location. Projection effects can be excluded as a reason for the opposite flow direction and polarity, as $\mu\!\gtrsim\!0.8$ and the tip of CEF-2 was located on the center side of the main sunspot of the group. The continuum images exhibit a continuous filamentary structure from the tip to the main body of CEF-2. The sign of the flow and magnetic field in this region is consistent with a downflow at the locality after being nearly vertical at the filament's head. The filament that harbored CEF-2 became more horizontal in the body and finally bent over to return into the Sun at the tail.  In that location within the tip, strong fields were observed. When CEF-2 moved away from the umbra, the magnetic field returned to nominal penumbral values.

\subsection{Expulsion of the CEFs}\label{sec:expulsion}

After CEF-1 and CEF-3 grew to occupy almost the entire distance from umbral to the outer penumbral boundary, the entire region containing the CEFs started to move. The temporal evolution of these regions harboring CEFs shows a radially outwards motion from the place they first appeared within the penumbra. They moved towards the outer boundary of the main sunspot of the group parallel to the penumbral filaments.  We can trace the location of the CEFs at all times, as the direction of $v_{\rm LOS}$ within them stayed opposite to the local normal Evershed flow of the surrounding penumbra. Hereafter, we refer to the outward motion of CEFs from the place they initially appear as their \textit{expulsion}.

 \begin{figure*}[phtb]
     \centering
     \includegraphics[width=1\textwidth]{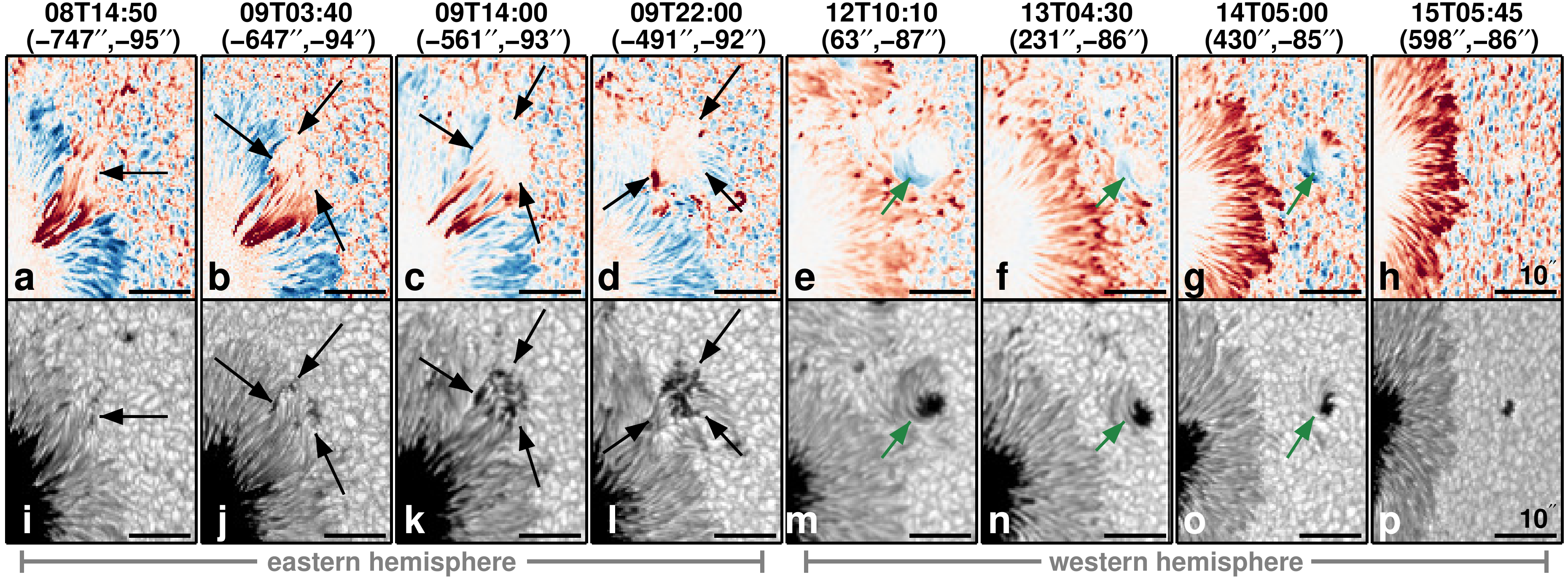}
     \caption{Maps of the $v_{\rm LOS}$ (top) and continuum (bottom) during the expulsion of CEF-1 and after it left the penumbra. $v_{\rm LOS}$ are clipped at $\pm$4\,\kms{}. In the first two columns, \ARA{} was located on the eastern solar hemisphere, while in the last four columns \ARA{} is on the western hemisphere. This change in viewing geometry between hemispheres of \ARA{} causes the normal Evershed flow to appear in panels (a) to (b) blueshifted and the CEF redshifted, while in panels (e) to (h) this pattern is reversed. The time and heliocentric coordinates of each scan are marked on the top part of each column. The full temporal evolution of CEF-1 is shown in the middle row of \animationone{} part of the online material. 
     }
     \label{fig:pore}
 \end{figure*}

We used the available low-cadence Hinode/SOT-SP scans for CEF-1 and the SDO/HMI data for CEF-3 to estimate the apparent velocity of the expulsion of the CEFs through their proper motion. The restriction of SDO/HMI is the low spectral resolution; however, SDO/HMI provides continuous 45\,s-cadence LOS velocity and magnetic field measurements, albeit at a single height (see \animationtwo{}). For the two data sets, we masked the CEFs and calculated the location of the center of gravity $R$ of a quantity $F$ within the CEF as

\begin{equation}
 R_{F}=\frac{\sum_{\{i,j\}\in A_{\rm CEF}} F_{ij}\sqrt{({r}_0-{r}_{ij})^2}}{\sum_{\{i,j\}\in A_{\rm CEF}} F_{ij}},  
\end{equation}

\noindent where $A_{\rm CEF}$ is the area covered by the CEF, $i,j$ identify pixels inside the CEF (identified using the $v_{\rm LOS}$ maps) and ${r}_0$ is the reference point chosen inside the closest umbra-penumbra boundary.  By replacing the placeholder $F$ by the parameters $I_{\rm c}$, $B$ or $v_{\rm LOS}$, we obtained the centers of gravity of the brightness ($R_{I_{\rm c}}$), of the magnetic field ($R_{B}$), and of the LOS velocity ($R_{v_{\rm LOS}}$). 

In Figure~\ref{fig:growth} we present the temporal evolution of $R_{v_{\rm LOS}}$ (black line), $R_{B}$ (dark gray), and $R_{I_{\rm c}}$ (light gray). The blue line shows the temporal evolution of the area of the CEFs. Before CEF-1 and CEF-3 were expelled, $R_{B}$ was closer to the umbra, while $R_{I_{\rm c}}$ is located in the mid-penumbra. This displacement between the centers of gravity comes from the fact that the field strength increases towards the umbra, also inside the CEFs. When these CEFs started moving the distance between the centers of gravity reduced until they coincide.

The horizontal velocity of expulsion for CEF-1 is on average $65$\,\ms{} (red line in Figure~\ref{fig:growth}(a)). This horizontal velocity traces the proper motion of the entire CEF-1 on the surface of the Sun, and not the plasma flow velocities within the penumbral filaments harbored inside CEF-1. The vertical dashed line marks the time when the magnetic structure that forms CEF-1 leaves the penumbra and a new spot starts forming. The maximum area of CEF-1 inside the penumbra is 24.7\,Mm$^2$. In addition, the decay of CEF-1 reveals that it is composed of individual strands harboring oppositely directed flows. While the center-of-gravity of CEF-1 moves smoothly radially outwards, an individual strand was observed moving with a speed ten times larger than the center-of-gravity velocity (see the middle row of \animationone{}).

We marked the expulsion of CEF-3 at five different moments (see vertical lines in Figure~\ref{fig:growth}(b)). The reference time is 2014 February 1 at 04:00\,UT, which corresponds to the first frame of \animationtwo{} provided as online material. CEF-3 was sporadic (i.e., it appeared and disappeared) in the early phase which lasted for 2 hours until it reached an area of $\sim$3\,Mm$^2$.  Figure~\ref{fig:growth} starts at this time. At $t_0=\text{10:05\,UT}$ the size of CEF-3 started to grow almost linearly in area at a rate of $130$\,km$^2$\,s$^{-1}$. Approximately five hours later ($t_1=\text{14:40\,UT}$), CEF-3 started to be expelled with a horizontal velocity of $117$\,\ms{}. Its maximum magnetic flux density and maximum $v_{\rm LOS}$ were reached at $t_2=\text{18:50\,UT}$, well before it reached its maximum area (13.1\,Mm$^2$) at $t_3=\text{20:58\,UT}$. This, again, is because the strongest fields and $v_{\rm LOS}$ values are to be found at or close to the umbral boundary. The innermost part of CEF-3 reached the outer penumbral boundary at $t_4=\text{02:30\,UT}$ on February 2. After $t_4$, a new spot started forming in the moat of the original host sunspot. The opposite-directed flow with respect to the adjacent penumbra inside the new spot suggests that this spot is formed from the same magnetic structure which previously formed CEF-3. The further evolution plotted in Figure~\ref{fig:growth} follows this spot.

\begin{figure*}
    \centering
    \includegraphics[width=.88\textwidth]{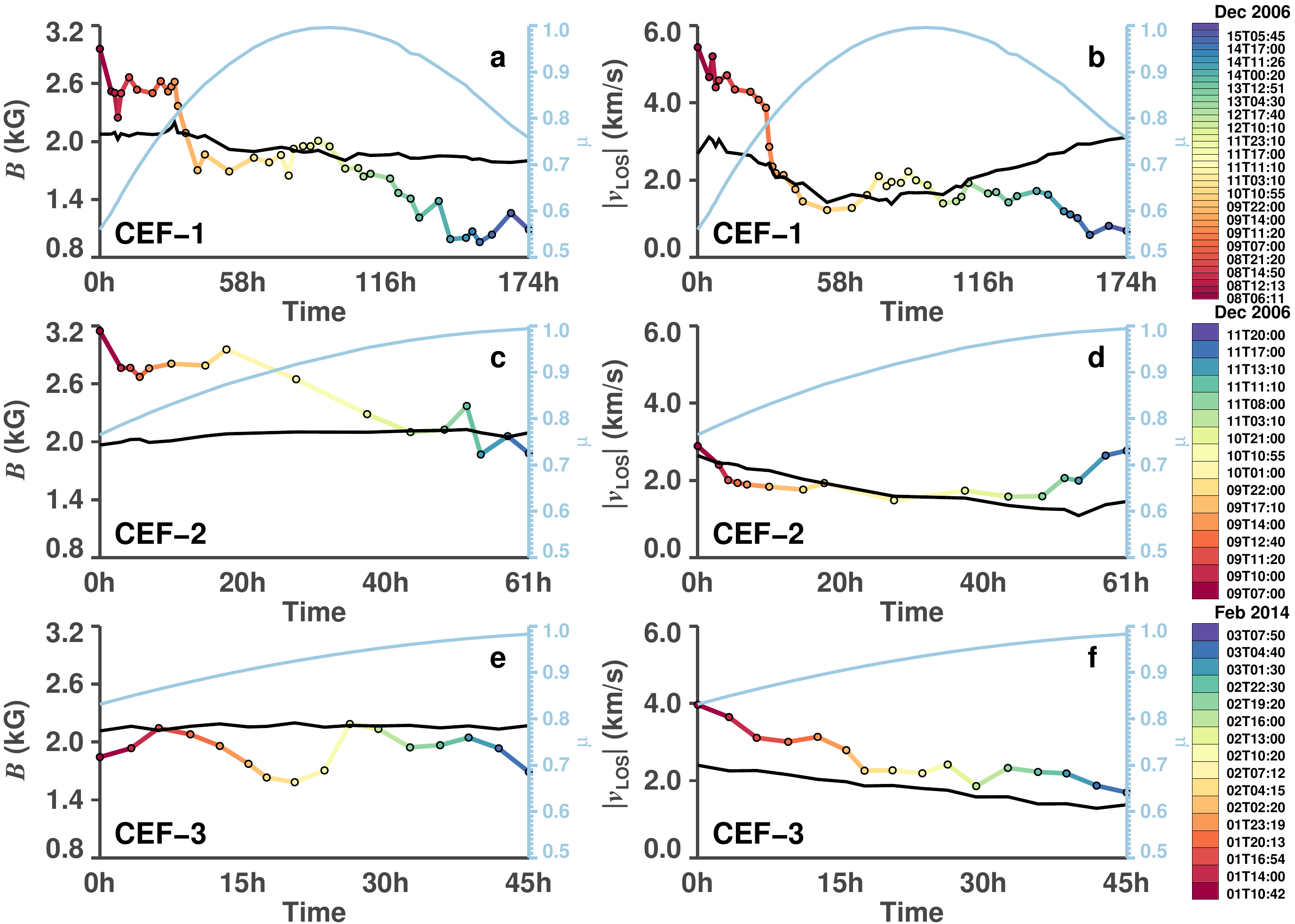}
    \caption{Temporal evolution of the magnetic field strength (left column) and $|v_{\rm LOS}|$ (right column) at $\tau_{5000\,\AA{}}=1$ inside the ROI in \ARA{} and \ARB{} and their CEFs. The black lines display the magnetic field strength and $|v_{\rm LOS}|$ averaged over the entire ROI inside the sunspot group. Colored lines show the mean values within CEF-1 (top), CEF-2 (middle) and CEF-3 (bottom), while the colors indicate the Hinode/SOT-SP scan times, starting from red and progressing to blue. The light-blue curve in each panel (referring to the right axis) indicates the $\mu-$values of the scans.}
    \label{fig:BVtemporal}
\end{figure*}

CEF-2 also undergoes dynamical changes and moves away from the region where it first appeared within \ARA{}. However, a different mechanism seems to be at work here. Recall that CEF-2 was located in between the north spot (main) and the south spot (satellite). On 2006 December 10, the satellite spot started to slowly rotate counterclockwise.  The temporal evolution suggests that CEF-2 followed the counterclockwise rotation of the satellite spot, indicating that it was anchored in the satellite umbra and was stretched by the satellite spot's rotation until CEF-2 disappeared (see e.g., Figure~\ref{fig:vlosAR10930}). This stretching of CEF-2 can be seen in the bottom panels of \animationone{} provided as online material.

\newpage
\subsection{What happens with the CEF magnetic structure after its expulsion?}\label{sec:expulsionCEFs:satelliteSpot}

During the expulsion of CEF-1, in the outer penumbra of the main spot of \ARA{}, or just outside its boundary, a number of pores developed, which then coalesced to form a small umbra with a penumbra attached to it (Figure~\ref{fig:pore}). Panel \ref{fig:pore}(i) shows the CEF-1 when it was located inside the penumbra of the main spot. In panel \ref{fig:pore}(j) four small pore-like dark regions appear (black arrows). These regions seem to merge and form a complex structure, as shown in panels \ref{fig:pore}(k) and \ref{fig:pore}(l). In panel \ref{fig:pore}(m)  the new feature has coalesced into an umbra that forms a penumbra on two sides including the one facing the penumbra of the main spot. The flow inside the newly formed penumbra has the same flow direction as the CEF-1 had when it was located inside the penumbra of the main spot. This flow pattern can be seen in the change from a redshifted patch when \ARA{} was on the eastern hemisphere (black arrows) to a blueshifted patch of flow on the western hemisphere (green arrows). From the perspective of the small umbra, the flow running along the newly formed penumbra has the direction of the normal Evershed flow.  

The newly formed region has the opposite magnetic polarity compared to the main spot. The polarity of the new spot could be unambiguously determined from the 19 Hinode/SOT-SP scans of \ARA{} that were taken close to disk center ($\mu\!>0.9$). The newly formed region also showed a slow counter-clockwise rotation in the moat of the main spot of \ARA{}. The penumbra of the newly formed spot reached its maximum area around 20\,UT on December 11, before it started to decay to a small pore on Dec 14. This pore was present for at least six days, before disappearing behind the west limb. The full temporal evolution can be seen in the middle row of \animationone{}, which is part of the online material. These observations suggest that the origin of the small spot is closely related to the magnetic structure that harbored CEF-1.

 \begin{figure*}[hbtp]
 \begin{center}
 \includegraphics[width=.95\textwidth]{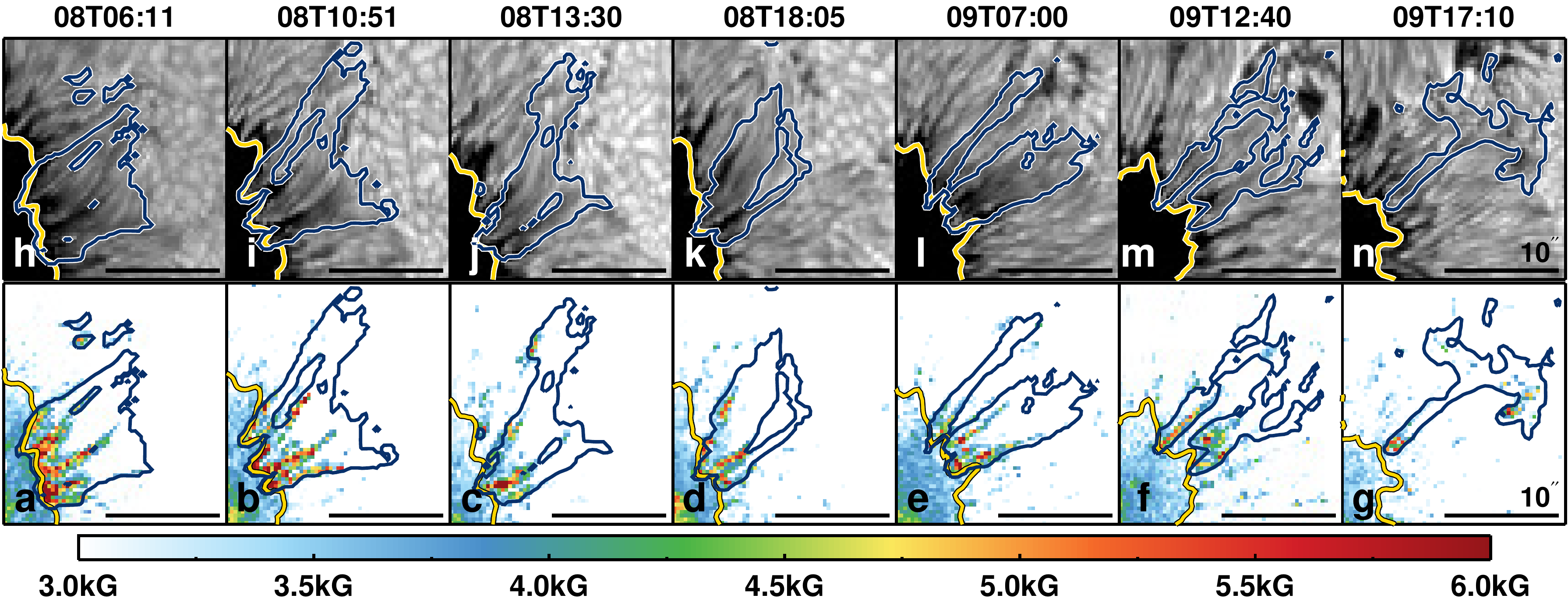}
 \caption{Location of the strong magnetic field in CEF-1. The two rows show the temperature (top) and magnetic field strength (bottom) at $\tau=1$. Contours mark the umbra-penumbra boundary (yellow) and CEF-1 (blue).} \label{fig:MapexpulsionB}
 \end{center}
 \end{figure*}

CEF-3 was expelled into a region where the penumbra appeared to be extended in a way that suggested a separate penumbra attached to the main penumbra of the spot (in particular it suggests the same polarity and curvature of the field, see \animationtwo{}). Once outside the main penumbra of the sunspot, CEF-3 appeared to form small patches of penumbra, moving radially outward from the host sunspot in the moat of \ARB{}, increasingly resembling an orphan penumbra (see \animationtwo{}).

 \subsection{$B$ and $v_{\rm LOS}$ inside the CEFs}\label{sec:magdistriExpulsion}

The magnetic field strength and $|v_{\rm LOS}|$ within the CEFs were taken from the spatially coupled inversion of the Hinode/SOT-SP data. The left and right columns of Figure~\ref{fig:BVtemporal} show the temporal evolution of the averaged $B$ and $|v_{\rm LOS}|$, respectively. The black lines are the averaged values within the region of interest (ROI) inside the sunspot group. We define the ROI as the full map displayed in Figures~\ref{fig:vlosAR10930} and \ref{fig:vlosAR11967}, where we masked out the quiet Sun (as the CEFs are present only in penumbrae), and the dark umbra (where the inversion results are less reliable due to the blends with molecular lines that form at low umbral temperatures). From top to bottom, the color-coded lines are the averaged values of $B$ and $|v_{\rm LOS}|$ for CEF-1 to CEF-3.  Color-coded marks indicate the different time stamps of the scans. For CEF-1 and CEF-2 some scans overlap because CEF-1 and -2 are partly present at the same time, hence some data points appear in both the top and middle panels. 

The mean $B$ value within the ROIs was on average $\sim$2\,kG, and showed little variation over the course of the evolution of the active regions. In the case of CEF-1, superstrong fields were observed in the early stages, when the area of CEF-1 was largest and filled the entire penumbral sector. In this phase, the CEF-1 reached the umbra-penumbra boundary (Figure~\ref{fig:BVtemporal}(a)). In a later stage, CEF-1 showed moderate field strengths, similar to the field strengths observed in CEF-2 and -3 although CEF-2 also harbored individual pixels with field strengths reaching 6\,kG.

For CEF-2 the strongest magnetic fields occurred at the time of its appearance, while the mean magnetic field of the spot was around $\sim$3.2\,kG (Figure~\ref{fig:BVtemporal}(c)). Magnetic fields larger than 4\,kG were seen inside CEF-2 during its formation phase. The mean $B$ remained at a high value until about 20\,hr after its appearance. Thereafter it decreased.
 
The magnetic evolution of CEF-3 is different compared to the other two CEFs. The mean value of $B$ inside CEF-3 oscillated around $\sim$1.9\,kG (Figure~\ref{fig:BVtemporal}(e)). The general trend of decreasing mean field strength with time, as seen for CEF-1 and CEF-2, is not visible in CEF-3.

The $v_{\rm LOS}$ values strongly depend on the projection ($\mu$-value), and therefore we do not compare their values one-to-one between different scans, but rather provide a qualitative description of their evolution. For scans observed close in time, the $\mu$-variation between scans is small, which allows us to describe roughly the temporal evolution of the line-of-sight velocity.

The temporal evolution of the line-of-sight velocity shows that  CEF-1 harbored considerably larger $|v_{\rm LOS}|$ values than the other two CEFs (Figure~\ref{fig:BVtemporal}, right column). Particularly during the early scans, CEF-1 was characterized by supersonic $|v_{\rm LOS}|$. The photospheric sound speed lies typically in the range $c_{s}\!\sim\!6\!-\!8$\,\kms{}. These fast $|v_{\rm LOS}|$ were co-temporal and roughly co-spatial with the superstrong magnetic fields found in CEF-1. In the late stages of CEF-1, the velocities returned to nominal penumbral values. CEF-2 and CEF-3 showed mainly low $|v_{\rm LOS}|$ values, with CEF-2 having a few points with clearly supersonic flows (roughly similar in number to points having $B>4$\,G).
 
The early superstrong fields in CEF-1 were located in the same pixels as those first reported by \citet{Siu-Tapia2017A&A, Siutapia2019}. These strong magnetic fields within CEF-1 stayed mostly close to the umbra-penumbra boundary at all times  (Figure~\ref{fig:MapexpulsionB}). The number of pixels with strong fields decreased along with their maximum field strength (Figure~\ref{fig:strongfields}) at the time when CEF-1 lost contact with the umbra. After the complete expulsion of CEF-1, the magnetic field strength, as well as the other atmospheric parameters in the patch of penumbra that had previously hosted it, returned to typical penumbral values (see e.g., Figure~\ref{fig:vlosAR10930}, \texttt{SCANS-A14}).
 
Figures~3.4 and 3.7 of \citet{CastellanosDuran2022...Phd} show the distributions of $B$ and $|v_{\rm LOS}|$ inside these three CEFs and how they vary over time. As discussed previously, those figures show a high number of pixels with strong magnetic fields and fast LOS velocities when CEF-1 and CEF-2 were in contact with the umbra-penumbra boundary. CEF-3 did not touch the umbra-penumbra boundary and strong magnetic fields on the side of CEF-3 that was closer the umbra-penumbra boundary were not present in CEF-3 at any time.

\begin{figure*}
    \centering
    \includegraphics[width=.99\textwidth]{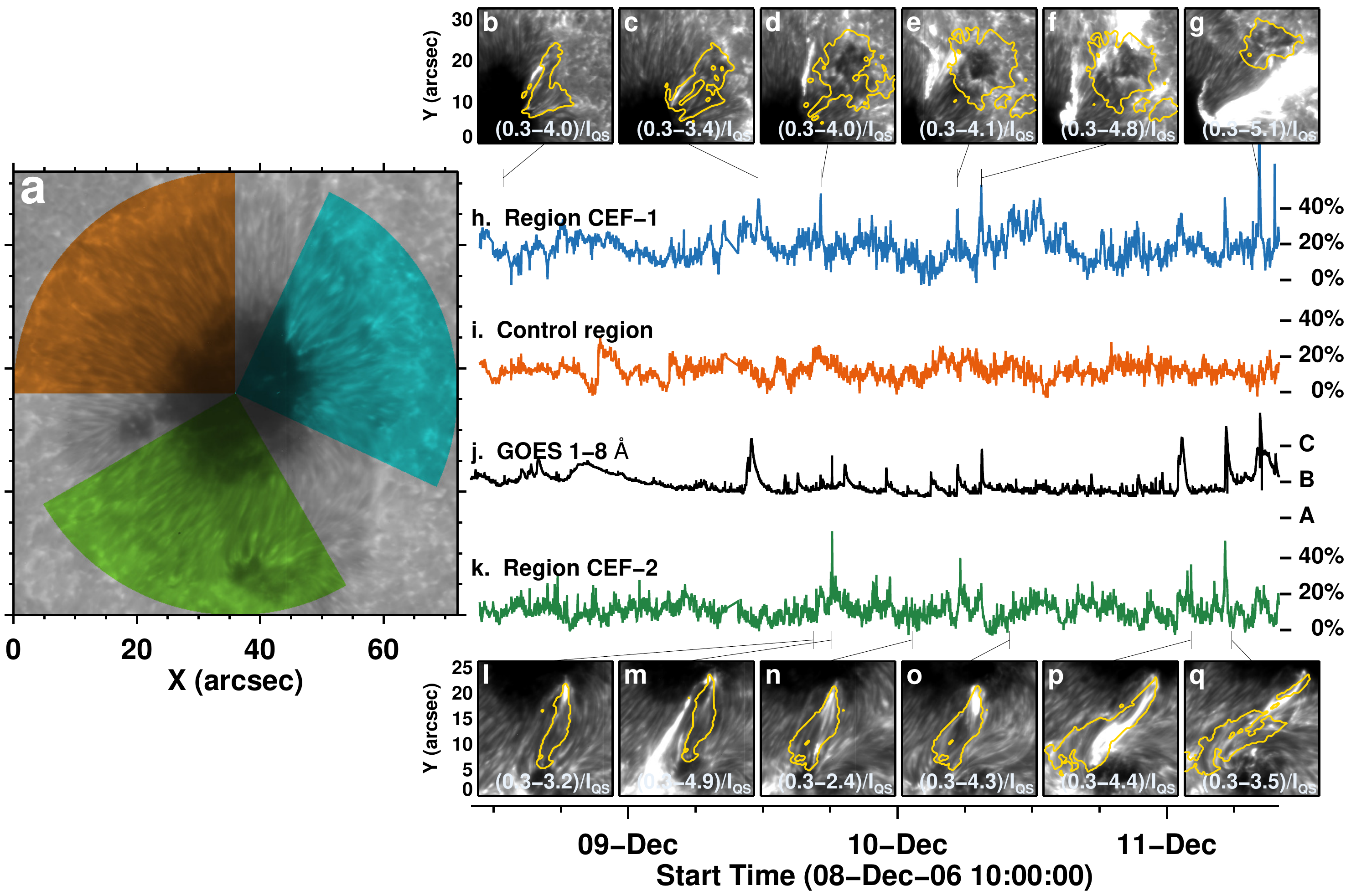}
    \caption{(a): Filtergram image of the chromospheric \ion{Ca}{2}\,H lines.
    (b)-(g): Examples of the chromospheric brightenings above CEF-1. (h), (i) and (k): Light curves of the \ion{Ca}{2}\,H mean intensity inside the three circular-sectors enclosing CEF-1 (blue), CEF-2 (green) and a control region without CEF (orange) marked in panel (a). See main text for how the circular sectors were selected. (j): GOES light curve at 1-8\,\AA{}. GOES classes A to C represent the X-ray flux integrated over the entire Sun in logarithmic scale ranging from $10^{-8}$ to $10^{-6}$\,W\,m$^{-2}$, respectively. Images (b)-(g) and (l)-(q) show examples of brightening events observed above CEF-1 ((b)-(g)) and CEF-2 ((l)-(q)). Locations of CEFs are marked with yellow contours. The time is marked with respect to the light curves in panels (h) and (k).  Images are normalized to the averaged quiet-Sun intensity and their dynamic ranges are shown on the bottom left of each panel. 
  }
    \label{fig:chromosphere}
\end{figure*}

\subsection{Chromospheric response above the CEFs}\label{sec:expulsitionCEF:Chrosmophere}

While the continuum images of the CEFs look very similar to the normal penumbra, the chromosphere above these structures is much more dynamic. The chromospheric images of the \ion{Ca}{2}\,H line taken by Hinode/BFI show brightening events that are co-spatial or appear at the boundaries of CEFs \citep[cf.][]{Louis2014A&A...CEF}. These brightening events were observed repeatedly. To quantify this chromospheric activity, we calculated the radiative flux in the \ion{Ca}{2}\,H line within three circular sectors for \ARA{} that hosted CEF-1 and CEF-2. The aperture of these sectors is 90$^{\circ}$ with a radius of 36\arcsec. We selected the areas to be of the same size for an unbiased comparison. The aperture and radius of the sectors were chosen to fully contain the CEFs during all phases of their evolution, covering also the strong elongation of CEF-2.  In addition, tests were performed by varying the aperture and radius of the circular sectors (not shown). The similarity of the results obtained suggests that the discussion below does not depend on the selection of the sectors.

Figure~\ref{fig:chromosphere} displays the temporal evolution within the three sectors color-coded blue for CEF-1, green for CEF-2, and orange for a control region containing only a typical penumbra region without any CEF. The three-light curves (h), (i), and (k) are normalized dividing by the area inside the circular sector and the averaged quiet-Sun intensity. Since we are interested to quantify the brightenings, i.e., short peaks in the light curve, rather than the long-term evolution of the sunspot group, we fitted the background with a 10$^{\mbox{th}}$-order polynomial and subtracted this fit from the light curve. We also included the GOES 1-8\,\AA{} flux showing the soft X-ray activity integrated over the entire solar disk. The light curves of the two CEF regions indeed showed enhanced chromospheric emission. Examples of associated brightenings appearing above or next to the location of CEFs range from small events (Figure~\ref{fig:chromosphere}(b), (c), (l), (n)) to a \textit{large} C-class flare seen in soft X-rays (Figure~\ref{fig:chromosphere}(g)).

\begin{figure}
    \centering
    \includegraphics[width=.45\textwidth]{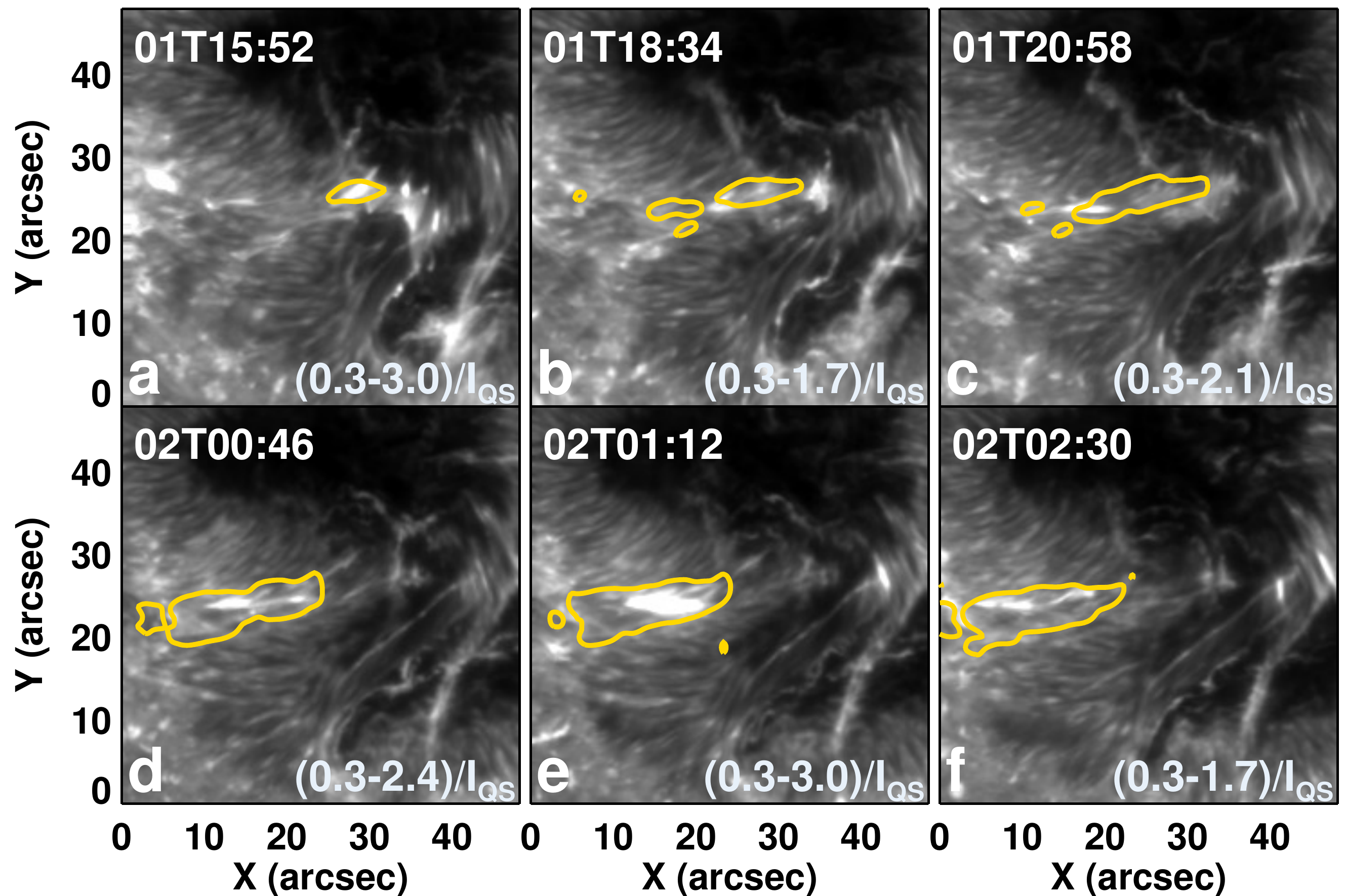}
    \caption{Examples of chromospheric brightenings observed in the \ion{Ca}{2}\,H line above CEF-3. The dynamic range of the images, normalized to the averaged quiet-Sun intensity, is given in the bottom part of each panel. Yellow contours mark the location of CEF-3 in the underlying photosphere.}
    \label{fig:chromospherecef3}
\end{figure}

A similar analysis was carried out for CEF-3. Little brightening events are also observed above CEF-3 (see Figure~\ref{fig:chromospherecef3}), however, their frequency and intensity are lower compared to the high chromospheric activity above CEF-1 and CEF-2. The complex magnetic topology of \ARB{} and the continuous chromospheric activity all over \ARB{} makes the chromospheric activity above CEF-3 only a minor contributor.

\section{Discussion}\label{sec:expulsionCEFs:discussion}

We analyzed the photospheric properties inside three CEFs using spatially coupled inversions. We also consider the  influence of CEFs to the chromosphere. We followed the temporal evolution of the CEFs by inverting all the available spectropolarimetric maps taken by Hinode/SOT-SP of the sunspot groups harboring them. The response at chromospheric heights above the CEFs was characterized using the filtergraph images in the \ion{Ca}{2}\,H line. Table~\ref{tab:properties_CEF} summarizes the properties of the three CEFs analyzed. We found that the CEFs are expelled from the location in the main sunspot of the group at a velocity of about $\sim$100\,\ms{} where they emerged radially outwards into the moat of the sunspot. To our knowledge there is just one report that showed the expulsion of a CEF \citep{Kleint2013ApJ}, however, that study focused on the so-called {\it umbral filaments} and did not provide further information about the CEF beyond the movement of the CEF. 

The analyzed CEFs appear to be the result of two different processes. Although there were no Hinode/SOT data available during the appearance phase of CEF-1, Hinode/SOT-BFI and -NFI images of December 6 and early 7, when \ARA{} appeared on the east limb and before CEF-1 was formed, were available. \citet{Siu-Tapia2017A&A} suggested that CEF-1 resulted from the coalescence of a satellite spot with the main spot, which inherited the penumbra of the satellite spot. For CEF-2 and CEF-3 it was possible to follow their entire formation process. These two CEFs appeared as intrusions within a fully developed penumbra without the merging with any external magnetic structure already visible on the solar surface. These intrusions mimic the appearance of new magnetic flux at the surface of the Sun. Similar emergence-like CEFs were observed in MHD simulations \citep{Chen2017ApJ...ARSimulation...CEF}.

Using MHD simulations, \citet{Siu-Tapia2018ApJ} proposed that CEFs can be driven by siphon flows. The gas pressure difference required to drive these flows can originate from any process that leads to a field strength enhancement at the endpoint of the flow. For a CEF, this is at the boundary between the umbra and penumbra. Such field strength enhancements were indeed observed for CEF-1 and CEF-2, making the siphon flow a possible driver of these two flows. However, for CEF-3, no such field strength enhancement was observed. 

CEFs showed a slightly different inclination relative to the surrounding penumbrae. This indicates that the reversed flow direction, which is the signature of CEFs, is associated by and likely driven by a somewhat different magnetic structure. Indeed, this is consistent with the finding of \citet{Siu-Tapia2018ApJ} that CEFs are driven by a siphon flow, while the normal Evershed flow is not.

\begin{table*}[t]
    \begin{center}
\begin{tabular}{l|ccc}
\toprule[1.5pt]
ID & CEF-1 & CEF-2 & CEF-3 \\
\midrule[1.5pt]
NOAA AR ID & 10930 & 10930 & 11967 \\
Date first observed& 2006 Dec 8 & 2006 Dec 7 & 2014 Feb 01\\
SP/\texttt{SCAN} ID & \texttt{A} & \texttt{A} & \texttt{B}  \\
Observed in SP/\texttt{SCANS} & 00-16 & 08-20 & 00-10 \\
SDO/HMI & No & No & Yes \\
Lifetime$^{a}$ (hr) & 50 & 49 & 24 \\
$\mu$ range & 0.56-0.92 & 0.76-0.98 & 0.83-0.96 \\
Maximum area$^{b}$ (Mm$^2$) & 24.7 & 12.3 & 17.7 \\
Opposite polarity$^{c}$? & No & Yes & Yes\\
Max $B(\log\tau=0)$ (kG) & 8.4 & 6.7 & 5 \\
Max $v_{\rm LOS}(\log\tau=0)$ (\kms{}) & 22.2 & 8.2 & 12.1 \\
Location strongest $B$ & UPB$^{d}$ & UPB$^{d}$ & EPPF$^{e}$\\
New spot formed? & Yes & No & Yes \\
Ejection mechanism & Radial$^e$  & Rotation AR & Radial$^e$ \\
\bottomrule[1.5pt]
\end{tabular}  
    \caption{Properties of the three expelled CEFs. $^a$Lifetime inside the penumbra. $^b$Maximum area of the CEF. $^c$Opposite polarity of the CEF with respect to the main spot. $^d$UPB: Umbra-penumbra boundary. $^e$EPPF: End-point penumbra filament. $^e$Radially outward towards the moat of the AR.
    \label{tab:properties_CEF}}
    \end{center}

\end{table*}
 
 CEF-1 (\animationone{}) and CEF-3 (\animationtwo{}) travelled radially outwards through the penumbra. When these CEFs reached the outer boundary of the penumbra of the main spot, a satellite spot started forming. The Evershed flow of this newly formed spot was originally the CEF of the main spot and did not change its flow direction (Figure~\ref{fig:pore}) when detaching from the main sunspot. This could suggest that the newly formed spot belonged to the same magnetic structure that formed the CEFs inside the penumbra of the main spot. The newly formed spot continued traveling into the moat of the main sunspot up to a distance similar to the length of the adjacent penumbra of the main sunspot. This distance coincides with the typical extension of the sunspot's moat, which often has the width of the penumbra in its vicinity \citep[e.g.,][]{Brickhouse1988SoPh..moat,Sobotka2007A&A...moat}. After the new spot travelled this distance, it stopped its radially outward motion. During this process, the new spot started decaying, losing its penumbra in the process (Figure~\ref{fig:pore}l).
 
There is evidence that connects the Evershed flow and the moat flow as its extension \citep{VargasDominguez2007ApJL...moatflow, VargasDominguez2008ApJ...moatflow, Rempel2011ApJ...Largescaleflows}. Also, there are previous reports of magnetic structures moving radially outwards from the penumbra into the moat, such as the expulsion of sea-serpent features and Evershed clouds \citep{Rimmele1994A&A...EF, SainzDalda2008A&AL...sea-serpent, CabreraSolana2006ApJL...EFclouds, CabreraSolana2007A&A...EC1, CabreraSolana2008A&A...EC2}. Sea-serpent features and Evershed clouds have (proper motion) expulsion speeds of $\sim$300--500\,\ms{}. These expulsion speeds are faster than the expulsion speeds of CEF-1 ($\sim$65\,\ms{}) and CEF-3 ($\sim$117\,\ms{}). The mean areas of sea-serpent features and Evershed clouds ($\sim$1.2--2.5\,Mm$^2$) tend to be smaller than the areas covered by CEF-1 ($\sim$10--25\,Mm$^2$) and CEF-3 ($\sim$2--13\,Mm$^2$). For all the features, the direction of the expulsion is parallel to the Evershed flow direction at this location. This suggests that the expulsion speed of a feature depends on its area, although the statistics are rather poor. We speculate that this may reflect a common mechanism responsible for the expulsion. This mechanism could be related to the Evershed flow itself, accelerating smaller features to higher velocities than larger ones. One possible test of this scenario would be to use the large sample of CEFs presented by \citet{CastellanosDuran2021...rareCEFs}. The sample covers a wide range of CEF areas. A common expulsion mechanism may show up in a correlation of the areas of CEFs with their expulsion speeds. 

The process leading to the expulsion of CEF-2 appears to be different from that affecting CEFs 1 and 3. The temporal evolution of CEF-2 suggests that its  disappearance is caused by the rotation of the satellite spot. CEF-2 was anchored in the satellite umbra and subsequently stretched by the satellite spot's rotation until it disappeared (Figure~\ref{fig:vlosAR10930}). Two studies found that the total rotation of the satellite spot in \ARA{} between 2006 December 10 and 2006 December 13 is $240^{\circ}-440^{\circ}$ \citep[][]{Zhang2007ApJL...CEF,Minoshima2009ApJ...flare}.  The rotation velocity of the spot increased almost linearly from $\sim$0.25$^{\circ}$\,hr$^{-1}$ to $\sim$8$^{\circ}$\,hr$^{-1}$  \citep[Figure\,8(c) of][]{Min2009SoPh} at the time when CEF-2 vanished.

CEF-1 and CEF-2 showed downflows co-spatial with strong magnetic fields. Strong $B$-values were always present at the umbra-penumbra boundary as long as CEF-1 was in contact with it. The area covered by strong magnetic fields and the maximum field strengths within these areas decreased when CEF-1 lost contact with the umbra. After the complete expulsion of CEF-1, the magnetic field strength and other atmospheric conditions in the same penumbral patch returned to normal. In the case of CEF-2, the gas flowing towards the main umbra was compressed by the strong field at the boundary of the umbra. The compression subsequently amplified $B$ and $v_{\rm LOS}$ to the observed high values in CEF-2. As with CEF-1, the magnetic field and $v_{\rm LOS}$ returned to nominal penumbral values after the expulsion of CEF-2 \citep[cf. Figures~3.5 and 3.7 of][]{CastellanosDuran2022...Phd}. The strong fields inside CEFs 1 and 2 could be related to the so-called ``magnetic barrier" \citep{vanNoort2013A&A} as proposed for CEF 1 by  \citet{Siutapia2019}. This mechanism was first proposed to explain the superstrong fields found at the endpoints of penumbral filaments. In the case of CEFs 1 and 2, the material flowing in a penumbral filament towards the umbra is forced to descend rapidly because of the presence of the strong umbral field acting as the magnetic barrier and hindering the flow from continuing. The magnetic barrier scenario also explains why $B$ and $v_{\rm LOS}$ returned to nominal values after the CEFs moved away from this barrier.

CEF-3 harbored strong fields of up to 5\,kG located at the endpoints of the penumbral filaments, similarly to the observations by \citet{vanNoort2013A&A}. Contrary to CEF-1 and CEF-2, CEF-3 emerged $\sim$1\arcsec{} away from the umbra-penumbra boundary. Therefore, no compression towards the umbra occurred there.  

In concordance with previous works \citep[e.g.,][]{Kleint2012ApJ...Cflare2CEF, Kleint2013ApJ, Louis2014A&A...CEF, Louis2020...cefs}, our data show many flares and brightenings associated with CEFs (Figures\,\ref{fig:chromosphere} and \ref{fig:chromospherecef3}). In addition, we also found increased chromospheric activity that appears to depend on how far the inner part of the CEF is located from the umbra-penumbra boundary. Thus, CEFs 1 and 2 that reach this boundary show considerably higher activity than CEF-3. 

The combination of the shear induced by the rotation of \ARA{} and the complexity of the  polarity inversion line (PIL) were proposed to be crucial to triggering the X3.4 flare \citep[SOL20061213T02:40; e.g.,][]{Kubo2007PASJ, Wang2008ApJAR10930, Schrijver2008ApJAR10930, Jing2008ApJ...676L..Xflare2CEF, Lim2010ApJ...719..Xflare2CEF, Gosain2010ApJ...Xflare2CEF, Fan2011ApJ...740...Xclassflare2CEF, Ravindra2011ApJ...Xflare2CEF, Inoue2011ApJ...Xflare2CEF, Inoue2012ApJ...760...Xflare2CEF, He2014JGRA..Xflare2CEF, Wang2022...XflareAR10930}. However, to our knowledge, previous studies neglected the opposite direction of the flow along the penumbral filaments at the location where the major flare was triggered. CEF-2 appeared in the middle of the penumbra and was then dragged/expelled with a rotation rate of  4$^{\circ}$\,hr$^{-1}$  \citep{Min2009SoPh} by the south satellite spot in \ARA{}. The remnants of CEF-2, visible in the $v_{\rm LOS}$ column of \animationone{}, coincide exactly with the location at the PIL which previous studies recognized as the region where this major flare was triggered. The presence of various opposite-directed flows, remnant from CEF-2 in this region, presents an extra factor in the complexity of the PIL and might therefore be another ingredient in triggering this X-class flare.

\section{Summary and conclusions}\label{sec:expulsionCEFs:conclusion}

In this study, we analyzed three CEFs observed in two sunspots groups. We investigate their temporal evolution, and their chromospheric impact.  In the following, we summarize the main results of our study:   

\begin{itemize}
    \item CEFs first appear close to or at the umbra-penumbra boundary and they grow until they reach the outer penumbral boundary.  
    
    \item Two different processes can explain the formation of the three CEFs part of this study. For CEF-1, \citet{Siu-Tapia2017A&A} suggested that it could have resulted from the coalescence of a satellite spot and the main umbra. Differently, CEF-2 and CEF-3 appeared as intrusions within a fully formed penumbra, independent of visible external magnetic structures \citep[cf.][]{Louis2014A&A...CEF, Louis2020...cefs,Guglielmino2017ApJ}. This behavior is compatible with the emergence of sub-surface magnetic flux within the penumbra. This was discussed for a simulated spot that is forming \citep{Chen2017ApJ...ARSimulation...CEF}. In these circumstances, CEFs are related to new flux (either emerging directly in the penumbra or just outside it). However, the CEFs studied here are within mature spots.
    
    \item  After a growth phase, CEFs 1 and 3 are seen to start moving parallel to the penumbral filaments. When they reach the outer part of the penumbra, a new spot starts forming in the moat of the main sunspot. The direction of the flow inside the penumbra of the newly formed spot is the same as in the CEFs and opposite to the adjacent penumbra of the main spot. This provides strong circumstantial evidence for a linkage between the CEFs and the newly formed spots. 
    
    \item In the moat, the newly formed spot reached a maximum distance to the penumbra at the outer boundary of the moat flow.
    
    \item  The expulsion speeds observed of CEF-1 and -3 in the penumbra are lower than  the ones of Evershed clouds \citep{CabreraSolana2006ApJL...EFclouds} and sea-serpent magnetic features \citep{SainzDalda2008A&AL...sea-serpent}. Considering CEFs are typically larger features (covering a larger area), one possible explanation is that these speeds depend on the size of the features. These photospheric features are often seen moving parallel to the penumbral filaments similar to CEF-1 and CEF-3. Common to all (CEFs, Evershed clouds, and sea-serpent features) is the presence of the normal Evershed flow surrounding these features and parallel to the direction of the expulsion. 

    \item \citet{Siu-Tapia2017A&A, Siutapia2019} showed for one Hinode/SOT-SP scan that superstrong $B$ observed in CEF-1 were associated with these flows directed towards the umbra, and that they were located mainly at the umbra-penumbra boundary. We confirm the presence of the superstrong fields in several Hinode/SOT-SP scans at different $\mu$-values. This makes a possible interpretation of a strongly Doppler shifted component as a magnetic component of strongly Zeeman splitted spectral lines less likely \citep[cf.][]{CastellanosDuran2020}. The temporal evolution of these superstrong $B$ showed that as soon as the expulsion of CEF-1 begins, and the contact to the umbra is lost, the maximum field strength drops. This supports the interpretation of \citet{Siutapia2019} that the origin of the superstrong fields in \ARA{} is related to compression at the magnetic barrier formed by the umbral field \citep{vanNoort2013A&A}.
    
    \item The expulsion mechanism of CEF-2 is influenced by the complex evolution of \ARA{}, and it is completely different from that of CEFs 1 and 3. CEF-2 was apparently dragged and subsequently stretched by the rotation of the satellite spot with a rotation rate of $\sim$4$^{\circ}$\,hr$^{-1}$.

\end{itemize}

Observers identify three physical processes that can lead to CEF formation: flux emergence \citep[e.g.,][]{Louis2014A&A...CEF, Louis2020...cefs}, adhesion of the penumbra from another spot after merging \citep{Siu-Tapia2017A&A}, and the association of granular and filamentary light bridges and CEFs \citep{CastellanosDuran2021...rareCEFs}. Further observations of CEFs and analyses of the deeper layers using simulated CEFs are needed to gain insight into the physical mechanisms responsible for their formation and maintenance. 

A total of 19 CEFs were identified in \ARB{}, however, in this study we focused on only two of them, for which multiple Hinode/SOT-SP observations were available. These 19 CEFs come on top of the 387 CEFs already reported by \citet{CastellanosDuran2021...rareCEFs}. An analysis of the known $\sim$400 CEFs could form the basis of an in-depth statistical study of CEF properties and evolution, to enhance not only our understanding of the nature of CEFs themselves, but also their impact on the sunspot dynamics and on the layers above CEFs.

In addition, the combination with new observations, in particular, stereoscopic observations between Hinode or SDO/HMI, combined with SO/PHI \citep{Solanki2020} onboard Solar Orbiter \citep{Muller2020A&A...SOLARORBITER}, will allow determining the two components of the velocity vector and not only the line-of-sight component. This will provide us with the necessary additional information to better understand CEFs.

\acknowledgments{We would like to thank the anonymous referee for careful reading and suggestions that improved the quality of the manuscript. J.~S. {Castellanos~Dur\'an} was funded by the Deutscher Akademischer Austauschdienst (DAAD) and the International Max Planck Research School (IMPRS) for Solar System Science at the University of G\"ottingen. This project has received funding from the European Research Council (ERC) under the European Union’s Horizon 2020 research and innovation program (grant agreement No. 695075). Hinode is a Japanese mission developed and launched by ISAS/JAXA, with NAOJ as domestic partner and NASA and UKSA as international partners. It is operated by these agencies in cooperation with ESA and NSC (Norway). SDO is a mission of the NASA's Living With a Star program.}

\section*{Appendix}

\setcounter{figure}{0}
\renewcommand{\thefigure}{A\arabic{figure}}

\begin{figure*}[tbhp]
 \begin{center}
 \includegraphics[width=.99\textwidth]{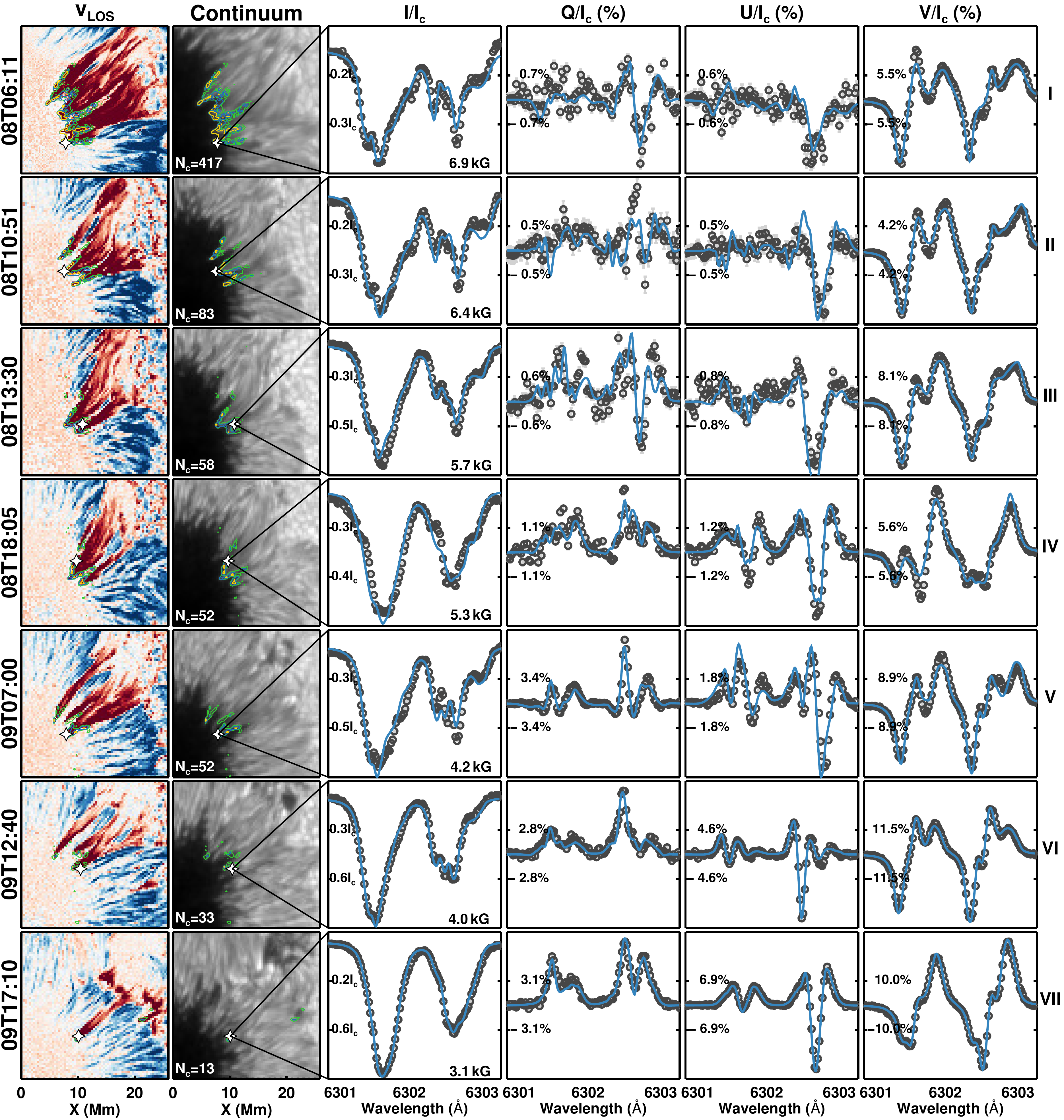}
 \caption{The full Stokes vector for seven different times in places where CEF-1 harbored superstrong fields. Time runs from top to bottom. The locations from where the profiles were extracted are marked on the $v_{\rm LOS}$ maps and continuum images displayed in the first two columns, respectively. The green, blue, and yellow contours on the same columns mark the $B(\log\tau=-0.8)$ at 3.5\,kG, 4\,kG, and 5\,kG levels, respectively. N$_{\rm c}$ is the number of pixels within the green contours. The observed Stokes profiles $I/I_{\rm c}$, $Q/I_{\rm c}$, $U/I_{\rm c}$, and $V/I_{\rm c}$ are displayed in columns 3--6 by the black circles, and the best fit to the data using the spatially coupled inversion is presented by the blue lines. The $B(\log\tau=-0.8)$ values at each location retrieved by the spatially coupled inversion are given in the panels in the third column (showing Stokes $I/I_{\rm c})$.  
}\label{fig:strongfields}
 \end{center}
 \end{figure*}

\paragraph{Figure~\ref{fig:strongfields}}  Examples of observed Stokes profiles and the location of the strong fields within CEF-1 as a function of time. Highly complex Stokes profiles were chosen to display the quality of the fits. The first two columns display the $v_{\rm LOS}$ and continuum maps of CEF-1, where CEF-1 can be identified as a red patch in the first column. Time runs from the top row to the bottom. Green, blue, and yellow contours on columns 1 and 2 mark the locations harboring fields stronger than 3.5\,kG, 4\,kG, and 5\,kG, respectively. The number of pixels inside the green contours (N$_{\rm c}$) are displayed in the second column. Columns 3 to 6 in Figure~\ref{fig:strongfields} show examples of observed Stokes profiles (gray open circles) and the fits using the spatially coupled inversions (blue lines). Notice that despite the high complexity of the observed Stokes profiles at these locations, the spatially coupled inversions obtained remarkably good fits \citep[e.g.,][]{CastellanosDuran2020}.

\paragraph{\animationone{}} Temporal evolution of CEF-1 and CEF-2 as seen by Hinode/SOT-SP.  This animation is composed of nine panels that mainly show the expulsion of CEF-1 and CEF-2. The columns display the continuum intensity, the magnetic field strength (clipped below 1\,kG and above 5\,kG), the LOS velocity (clipped between $\pm$3\,\kms{}), and the LOS inclination of the magnetic field. The top row shows the full \ARA{}, while the second and third rows present close-ups of CEF-1 and CEF-2 (black arrows). The cadence of the animation varies depending on the availability of the Hinode/SOT-SP maps. The first frame starts on 2006 December 08 at 6:11\,UT when \ARA{} was located in the solar western hemisphere at (--697\arcsec{}, --83\arcsec{}). The last frame was taken on 2006 December 15 at 13:02\,UT when \ARA{} was located in the eastern hemisphere at (711\arcsec{}, --86\arcsec{}). The duration of the animation is 5 seconds.

\paragraph{\animationtwo{}} Temporal evolution of CEF-3 as seen by SDO/HMI. The animation consists of four panels that show the expulsion of CEF-3. Panels display the continuum intensity (a),  LOS magnetic field (b), LOS velocity (c), and the location of CEF-3 (d; enclosed by black contours). The field of view covers an area of $\sim50\times50$\,Mm.  Thin contours in all panels mark the locations of the penumbra and umbra boundaries. The first frame starts on 2014 February 2 at 00:00\,UT when the sunspot group was at (540\arcsec{}, --130\arcsec{}). The last frame was taken on 2014 February 2 at 13:30\,UT when \ARB{} was at (--220\arcsec{}, --125\arcsec{}). The cadence between images is 45 seconds. The duration of the animation is 27 seconds. For better visibility of the processes in the penumbra, we masked out umbral pixels in panels (b) and (c).

\bibliographystyle{aa}
\bibliography{references}

\end{document}